\documentclass[%
 reprint,
superscriptaddress,
longbibliography,
 amsmath,amssymb,
 aps,
prb
]{revtex4-2}

\usepackage[T1]{fontenc}
\usepackage{graphicx}% Include figure files
\usepackage{dcolumn}% Align table columns on decimal point
\usepackage{bm}% bold math
\usepackage{amsmath}
\usepackage{lmodern}
\usepackage{mathtools}

\usepackage{xcolor}
\usepackage{graphicx}% Include figure files
\usepackage{dcolumn}% Align table columns on decimal point
\usepackage[hidelinks]{hyperref}% add hypertext capabilities

\usepackage[normalem]{ulem}%do przekreslen \sout{Hello World}

\hypersetup{
    colorlinks,
    linkcolor= [rgb]{0.1804,0.1882,0.5725},
    citecolor=[rgb]{0.1804,0.1882,0.5725},
    urlcolor=[rgb]{0.1804,0.1882,0.5725}
}

\usepackage[T1]{fontenc}
\usepackage{graphicx}
\usepackage{dcolumn}
\usepackage{bm}
\usepackage{amsmath}
\usepackage{lmodern}
\usepackage{mathtools}

\usepackage{upgreek}
\usepackage{chemformula}
\usepackage{tabularx}
\usepackage{makecell}

\usepackage{xcolor}
\usepackage{siunitx}
\usepackage{dcolumn}

\usepackage{tcolorbox}%boxes \tcbox[sharp corners, colframe=black, colback=white]{}
\usepackage{colortbl}

\usepackage{float}
\usepackage{soul}

\begin{document}

\title{Observation of second order meron polarisation textures in optical microcavities}

\author{M.\,Kr\'ol}
\affiliation{Institute of Experimental Physics, Faculty of Physics, University of Warsaw, ul.~Pasteura 5, PL-02-093 Warsaw, Poland}
\author{H.\,Sigurdsson}
\affiliation{Department of Physics and Astronomy, University of Southampton, Southampton, SO17 1BJ, United Kingdom}
\affiliation{Skolkovo Institute of Science and Technology, Moscow, Russian Federation}
\author{K.\,Rechci\'nska}
\author{P.\,Oliwa}
\author{K.\,Tyszka}
\affiliation{Institute of Experimental Physics, Faculty of Physics, University of Warsaw, ul.~Pasteura 5, PL-02-093 Warsaw, Poland}
\author{W.\,Bardyszewski}
\affiliation{Institute of Theoretical Physics, Faculty of Physics, University of Warsaw, Poland}
\author{A.\,Opala}
\affiliation{Institute of Physics, Polish Academy of Sciences, al.\,Lotnik\'{o}w 32/46, PL-02-668 Warsaw, Poland}
\author{M.\,Matuszewski}
\affiliation{Institute of Physics, Polish Academy of Sciences, al.\,Lotnik\'{o}w 32/46, PL-02-668 Warsaw, Poland}
\author{P.\,Morawiak}
\affiliation{Institute of Applied Physics, Military University of Technology, Warsaw, Poland}
\author{R.\,Mazur}
\affiliation{Institute of Applied Physics, Military University of Technology, Warsaw, Poland}
\author{W.\,Piecek}
\affiliation{Institute of Applied Physics, Military University of Technology, Warsaw, Poland}
\author{P.\,Kula}
\affiliation{Institute of Chemistry, Military University of Technology, Warsaw, Poland}
\author{P.\,G.\,Lagoudakis}
\affiliation{Skolkovo Institute of Science and Technology, Moscow, Russian Federation}
\affiliation{Department of Physics and Astronomy, University of Southampton, Southampton, SO17 1BJ, United Kingdom}
\author{B.\,Pi\k{e}tka}%\k{e}
\author{J.\,Szczytko}
\email{Jacek.Szczytko@fuw.edu.pl}
\affiliation{Institute of Experimental Physics, Faculty of Physics, University of Warsaw, ul.~Pasteura 5, PL-02-093 Warsaw, Poland}

\begin{abstract}
Multicomponent Bose-Einstein condensates, quantum Hall systems, and chiral magnetic materials display twists and knots in the continuous symmetries of their order parameter, known as Skyrmions. Originally discovered as solutions to the nonlinear sigma model in quantum field theory, these vectorial excitations are quantified by a topological winding number dictating their interactions and global properties of the host system. Here, we report the first experimental observation of a stable individual second order meron, and antimeron, appearing in an electromagnetic field. These complex textures are realised by confining light into a liquid-crystal filled cavity which, through its anisotropic refractive index, provides an adjustable artificial photonic gauge field which couples the cavity photons motion to its polarisation resulting in formation of these fundamental vectorial vortex states of light. Our observations take a step towards bringing topologically robust room-temperature optical vector textures into the field of photonic information processing and storage. 
\end{abstract}

\maketitle

Twists in the SO(3) order parameter of magnetic systems lead to topologically protected excitations known as skyrmions which are characterised by nontrivial spin textures~\cite{Nagaosa_NatMater2013, Wiesendanger_NatRevMater2016, Krause_NatMat2016, Fert_NatRevMater2017}. Just like quantised singular vortices in superfluid Helium or Bose-Einstein condensates these skyrmionic excitations are topologically robust against external perturbation since they cannot smoothly relax into the defect free ground state of the system, thus becoming highly important to understand phase transitions and critical behaviour in ordered many-body systems down to the quantum level~\cite{Lohani_PRX2019}. This robustness has also led to innovative proposals in the field of spintronics of stable information storage and processing with skyrmions at unprecedented spatial scales. They have been observed in chiral magnets~\cite{Muhlbauer_Science2009}, non-centrosymmetric magnets~\cite{Peng_NatNano2020}, surface plasmons~\cite{Du_NatPhys2019, Davis_Science2020}, exciton-polaritons~\cite{Donati_PNAS2016} to name a few, and have reached room temperature conditions in magnetic thin films~\cite{Yu_NatMat2011, Woo_NatMat2016} and liquid crystals~\cite{Ackerman_NatComm2017, Foster_NatPhys2019}.

\begin{figure*}[ht]
\centering
\includegraphics[width=\textwidth]{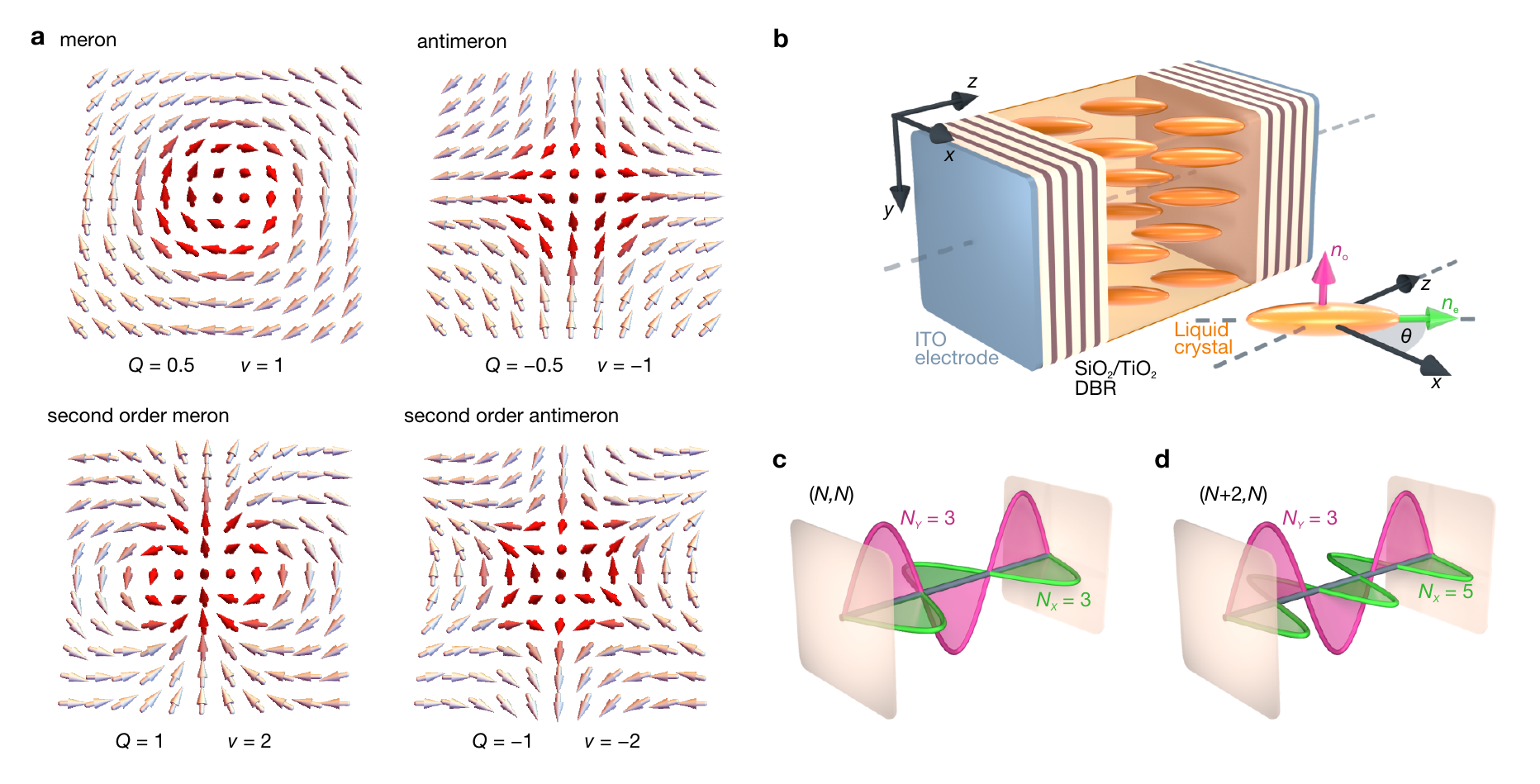}
\caption{\textbf{Meron polarisation textures.} \textbf{a}, Illustration of meron, antimeron, second order meron and second order antimeron textures. The arrows represent the order parameter $\mathbf{S}$ from equation~\eqref{eq:1}. \textbf{b}, Schematic of a microcavity filled with liquid crystal media. The liquid crystal microcavity can be tuned to contain perpendicularly polarised, degenerate modes \textbf{c} with the same mode numbers or \textbf{d} with different mode numbers. The electric field distribution of $X$ ($Y$) polarized mode is plotted in green (pink) colour.}
\label{im:Fig1}
\end{figure*}

Skyrmion textures appear as natural excitations in multicomponent quantum systems since a surjective homomorphism links the SU(2) unitary symmetry group to the SO(3) rotational symmetry group. In a photonic system, the two orthogonal polarisation components of the electromagnetic field can be described by a three-dimensional Stokes (pseudospin) vector located on the surface of the Poincar\'{e} sphere. Therefore, such topological knots and twists in an electromagnetic field can in-principle exist in the same sense as skyrmions in thin-film magnetic materials. Of special interest are spin textures known as magnetic vortices or ``merons'' which originate from Yang-Mills theory~\cite{Shifman_WoSc1994}. Due to their similarity to skyrmions they are sometimes referred to as half skyrmions or baby skyrmions since they can possess half of the skyrmion topological integer charge $Q$ defined in two-dimensional system as
\begin{equation} \label{eq:1} 
Q = \frac{1}{4\pi} \int \mathbf{S} \cdot \left( \partial_x \mathbf{S} \times \partial_y \mathbf{S} \right) dxdy,
\end{equation}
 where $\mathbf{S}$ is the order parameter. Alternatively, the charge of the meron can be determined through $Q = vp/2$ from its vorticity ($v$) and polarity ($p$) which describe the in-plane and out-of-plane order parameter orientation respectively~\cite{Senthil_Science2004}. The simplest configurations are those composed of $v=\pm1$ and $p=\pm1$ referred to as ``merons'' ($Q = 1/2$) and ``antimerons'' ($Q= -1/2$) (Fig.\,\ref{im:Fig1}a). 
In fact, twists in the Hamiltonian parameter space can be regarded as merons whose textures relate to the Berry curvature and charge determines the topological character of condensed matter systems~\cite{Bernevig_Science2006, Bleu_PRL2018, Gianfrate_Nature2020, Guo_PRL2020}. Interestingly, merons cannot exist as isolated objects unless spatially constrained~\cite{Shinjo_Science2000, Phatak_PRL2012}. They either form in lattices~\cite{Flayac_PRL2013, Vishnevsky_PRL2013, Yu_NatCommun2014, Yu_Nature2018, Cilibrizzi_PRB2016, Peng_NanoLett2017, Nych_NatPhys2017} or as paired objects only observed before in magnetic thin films~\cite{Gao_NatCommun2019}.

There also exist higher order merons, with vorticity $v=\pm2$ which are referred as ``second order merons'' ($Q = 1$) and "second order anti-merons'' ($Q = -1$) (Fig.\,\ref{im:Fig1}a). To the best of our knowledge, these second order twists in order parameter have not been observed in any system to date.

In this study, we present experimental and numerical evidence of second order merons 
and antimerons  
in the photonic field of an optical microcavity filled with a liquid crystal (LC) at room temperature. The second order merons appear as the natural eigenmodes of the system due to its tunable optical anisotropic structure. We demonstrate that a pattern of merons (antimerons) can smoothly merge to form a second order meron (antimeron). Effective Hamiltonians describing the two distinct meron textures are derived linking our observations to alternative low-dimensional condensed matter systems and paving the way towards synthesising fundamental order parameter twists on nonlinear optical fluids in the strong light-matter coupling regime.

We investigated microcavities with a birefringent LC layer enclosed between two parallel distributed Bragg reflectors (DBRs), as schematically shown in Fig.\,\ref{im:Fig1}b. The birefringent medium is characterised with two refractive indices: 
extraordinary $n_\textrm{e}$, parallel to the director of the LC molecules defining the long optical axis, and ordinary $n_\textrm{o}$, perpendicular to the director.  
This molecular director can be altered by application of external bias to transparent indium tin oxide (ITO) electrodes on the sample. We investigated a configuration in which the director rotates in the $x$--$z$ plane with applied field.
Different effective refractive indices $n$ for linearly polarised light along $x$ and $y$ axes lead to splitting of the optical modes  fulfilling standing wave condition for an optical path length $n d = N \lambda/2$ along the width of the cavity $d$, for incident wavelength $\lambda$ and mode number $N$. In a sufficiently wide cavity multiple optical modes with different mode numbers can be confined.  The unique property of LC-filled microcavity is the control over the energies of linearly $x$-polarised optical modes ($X$) with respect to $y$-polarised modes ($Y$) which allows to tune them in-and-out of resonance with respect to each other. In this work we concentrate on two different regimes where both $X$ and $Y$ modes have the same parity corresponding to $(N_X, N_Y) = (N,N)$ and $(N_X, N_Y) = (N+2,N)$ (see Figs.\,\ref{im:Fig1}c,d) which possess uniquely different photonic spin-orbit coupling mechanisms leading to meron and antimeron textures.

\begin{figure}[ht]
\includegraphics[width=.495\textwidth]{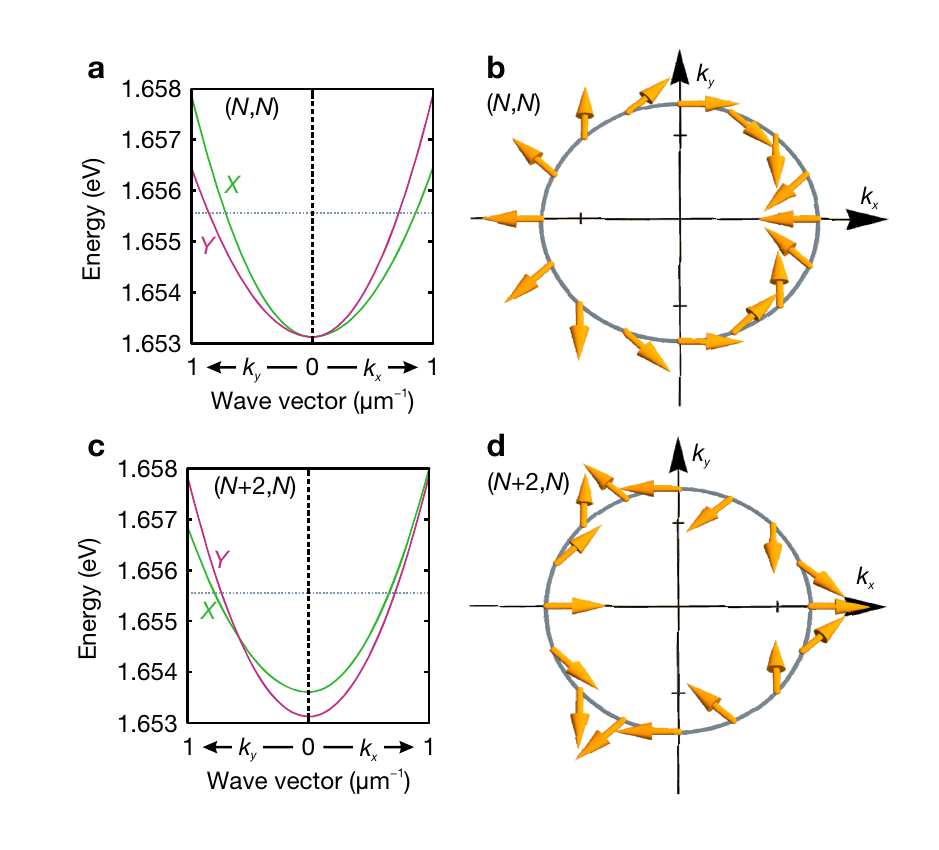}

\caption{\textbf{Spin structure in momentum space.}
\textbf{a}, Dispersion relation of cavity modes in $(N,N)$ regime along wave vector in $x$ ($k_x$) and $y$ ($k_y$) directions. \textbf{b}, Polarisation at constant energy cross section marked by dashed horizontal line in \textbf{a}. Polarisation state of the inner cavity mode is represented by the Stokes vectors $\mathbf{S} = (S_1,S_2,S_3)$ (yellow arrows). \textbf{c}, Dispersion relation of cavity modes in $(N+2,N)$ regime along $k_x$ and $k_y$. \textbf{d}, Polarisation of the inner cavity mode at constant energy marked by dashed horizontal line in \textbf{c} shown by Stokes vectors $\mathbf{S}$.}
\label{im:SIdisp}
\end{figure}

The optical eigenmodes, in the $XY$ polarisation basis, can be described by the following Hamiltonian with similar structure to the one describing TE-TM splitting in optically isotropic semiconductor microcavities~\cite{Kavokin_PRL2005}:
\begin{equation} \label{eq:2} 
  \hat{H} =  \epsilon(\mathbf{k})
- \left[\delta_x k_x^2 - \delta_y k_y^2 - \Delta E\right]\hat\sigma_z - \delta_{xy} k_xk_y\hat\sigma_x ,
%  E_k = \frac{\hbar^2 k_x^2}{2m_x} + \frac{\hbar^2 k_y^2}{2m_y}
\end{equation} 
where $\epsilon(\mathbf{k})=\hbar^2(k_x^2/m_x + k_y^2/m_y)/2$ describes cavity photons with masses $m_{x,y}$ along the $x,y$ direction respectively, $\delta_x, \delta_y, \delta_{xy}$ are parameters proportional to the birefringence $\Delta n=n_e-n_o$~\cite{Rechcinska_Science2019}, $\hat\sigma_{x,y,z}$ are the Pauli matrices, and $\Delta E=E_{Y,N_Y}-E_{X,N_X}$ is the $XY$ mode splitting at normal incidence ($k=0$). Notably, this splitting is equivalent to the presence of an effective magnetic field (Zeeman splitting) which plays the role of an artificial photonic gauge field applied to the structure.  In this sense, the polarisation of the cavity photons plays the same role as a two-level spinor for massive particles. The derivation of equation~\eqref{eq:2} from a simplified model of an optical two-dimensional waveguide filled with anisotropic dielectric medium is presented in the Supplementary Information. 

In the $(N,N)$ regime the molecular director is oriented along the $z$ axis, so $\theta=90^{\circ}$, and the refractive indices of the cavity medium are the same for normal-incident light polarised along $x$ or $y$ axis (i.e, $m_x = m_y$). Here, we have $\delta_x = \delta_y= \delta_{xy} > 0$ and $\Delta E = 0$ which gives rise to the standard optical spin Hall effect (see Figs.~\ref{im:SIdisp}a,b) observed for microcavity exciton-polaritons and bare cavity photons~\cite{Leyder_NatPhys2007, Maragkou_OptLett2011,Lekenta2018}. This unique interplay between the photons motion and polarisation results in a spatial polarisation texture composed of a meron-antimeron lattice, as previously observed in a microcavity exciton-polariton condensate~\cite{Cilibrizzi_PRB2016}.

On the other hand, the $(N+2,N)$ regime is obtained by changing the molecular director $\theta < 90^\circ$ which tunes the refractive index of cavity for light polarised along $x$ axis (see Fig.~\ref{im:Fig1}). 
In this regime one has $m_x \neq m_y$, $\delta_x, \delta_{xy}>0$, $\delta_y<0$. By detuning the modes slightly, $\Delta E < 0$, leads to severely different artificial spin-orbit coupling of the cavity photons (see Figs.~\ref{im:SIdisp}c,d).

\begin{figure*}[ht]
\centering

\includegraphics[width=\textwidth]{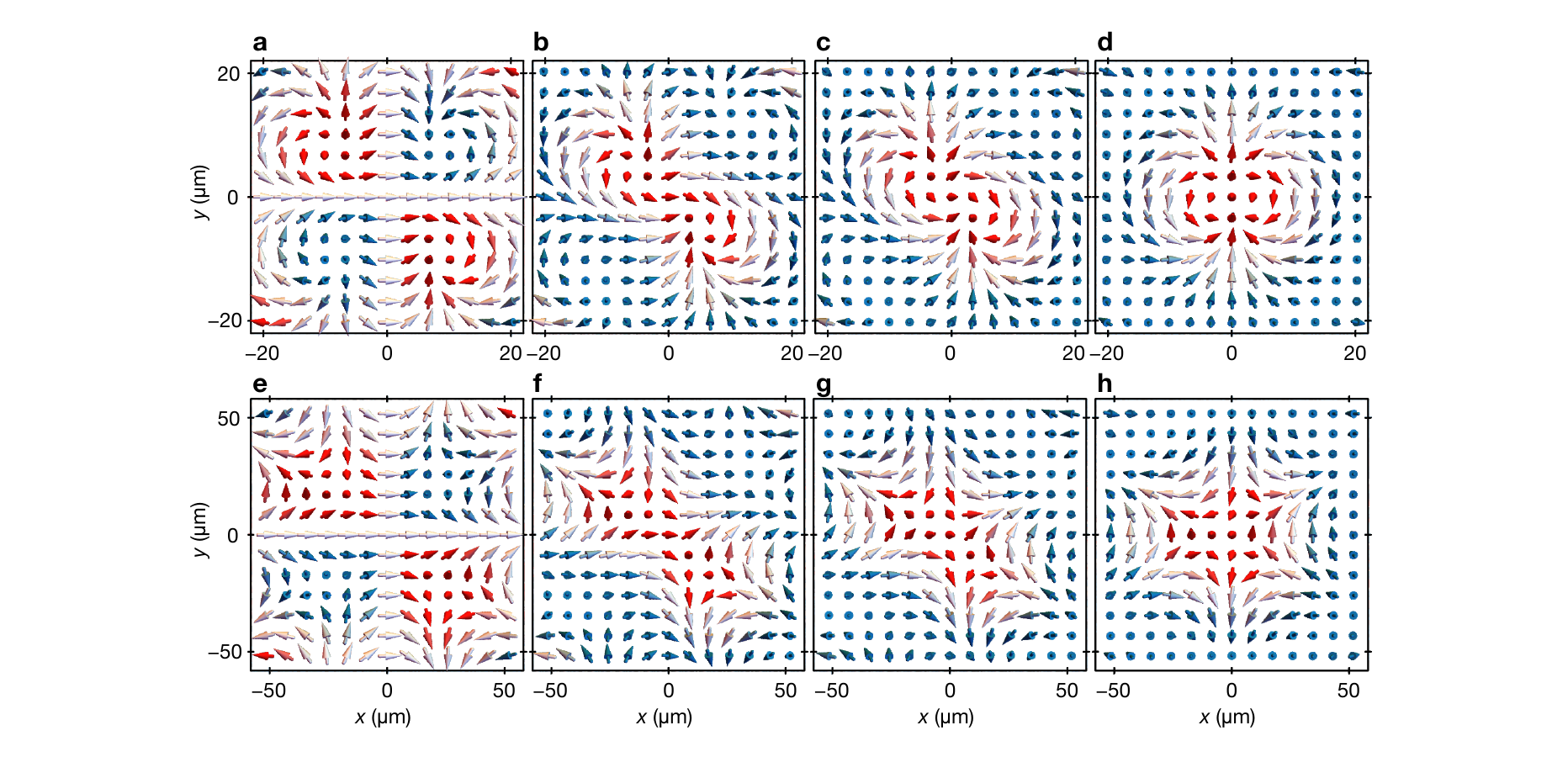}
\caption{\textbf{Spatial meron and antimeron polarisation textures in LC microcavities.} \textbf{a}--\textbf{d}, Berreman simulations for cavity in $(N,N)$ regime where the arrows represent the Stokes vector $\mathbf{S} = (S_1,S_2,S_3)$ and are coloured using the $S_3$ parameter. The excitation polarisation changes smoothly from linear (\textbf{a}) to circular (\textbf{d}) polarisation resulting in merons merging into a second order meron. \textbf{e}--\textbf{h}, Berreman simulations for cavity in $(N+2,N)$ regime. The same effect is observed with now two antimerons merging instead to form a second order antimeron as the exctiation polarisation changes from linear to circular.}
\label{im:Fig2big}
\end{figure*}

\begin{figure*}[ht]
\center%flexibility ing
\includegraphics[width=\textwidth]{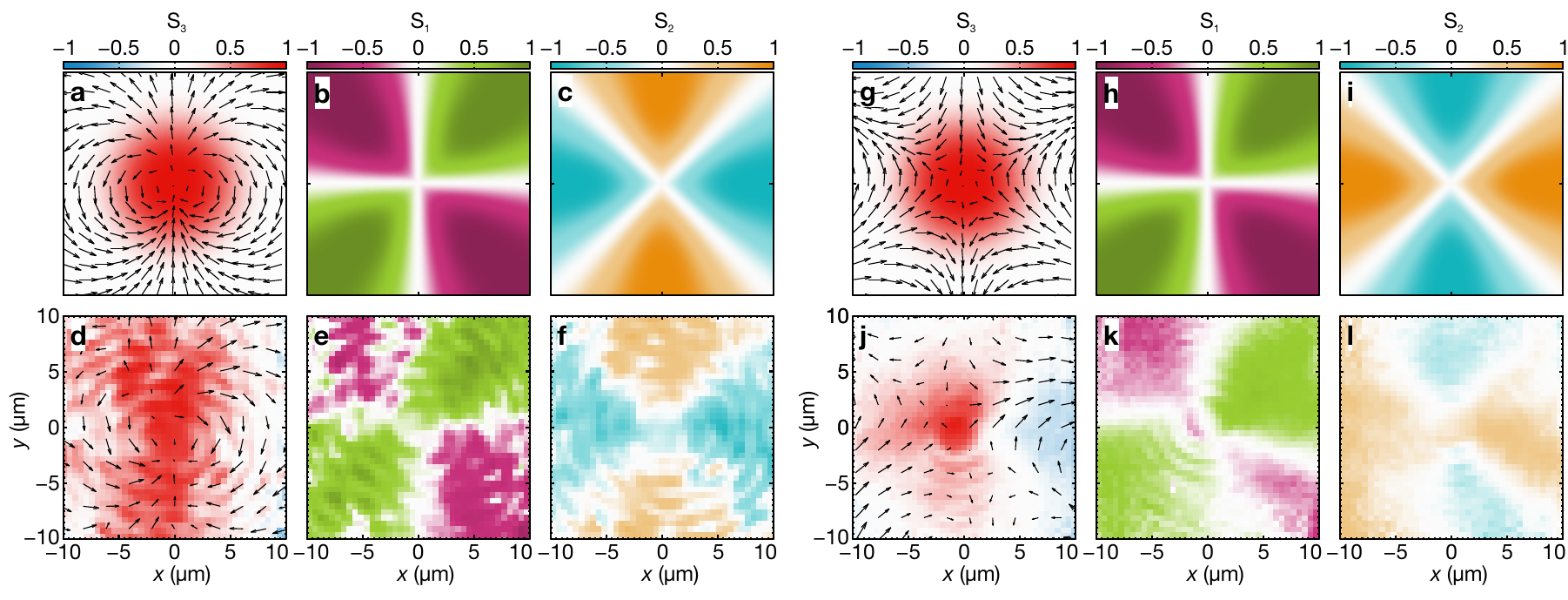}
\caption{\textbf{Second order meron and antimeron textures in LC microcavities.} \textbf{a}--\textbf{c}, $S_3, \ S_1$, and $S_2$ Stokes parameters showing the analytical spin texture of a second order meron given by equation~\eqref{eq:3}. Black arrows correspond to $\mathbf{S}_\parallel = (S_1,S_2)$. \textbf{d}--\textbf{f}, Experimental spatial polarisation texture of $\sigma^+$ polarised light transmitted through a LC microcavity in $(N,N)$ regime. 
\textbf{g}-\textbf{i}, $S_3, \ S_1$, and $S_2$ Stokes parameters showing the analytical spin texture of a second order antimeron given by equation~\eqref{eq:3}. \textbf{j}--\textbf{l}, Experimental spatial polarisation texture of $\sigma^+$ polarised light transmitted through a LC microcavity in $(N+2,N)$ regime.}
\label{im:Fig_Exp}
\end{figure*}

To illustrate the difference between the two regimes we show in Fig.~\ref{im:Fig2big} the real-space polarisation textures of light transmitted through a LC microcavity, calculated using the Berreman method~\cite{Berreman_1972} (see Methods). Figures~\ref{im:Fig2big}a-d show the adiabatic evolution of the polarisation texture in the $(N,N)$ regime for an excitation polarisation going from linear to circular. Figure~\ref{im:Fig2big}a shows the previously reported half-skyrmion lattice~\cite{Cilibrizzi_PRB2016}. Here, four merons of charge $Q=\pm 1/2$ -- two with positive and two with negative polarity -- can be observed in the quadrants of the system. When polarisation of the excitation beam changes to elliptical, two merons start merging and create a single second-order meron (also referred to as bimeron) of $Q=+1$ as the laser excitation becomes fully circularly polarised. 

In the $(N+2,N)$ regime, corresponding to Fig.~\ref{im:Fig2big}e-h, a very different behaviour is observed. Starting with linearly polarised incident light (Fig.~\ref{im:Fig2big}e) we observe again four antimerons with the same polarity but inverse vorticity compared to the $(N,N)$ regime. When the excitation polarisation is gradually changed to circular, two of these antimerons merge creating one second order antimeron $Q=-1$ presented in Fig.\,\ref{im:Fig2big}h.

This dramatic change in the topological integer charge $Q$ of these spin textures is precisely captured by equation~\eqref{eq:2}. The charge $Q$ has a different sign between the $(N,N)$ and $(N+2,N)$ regimes because of the polarisation structure of parabolic eigenmodes in momentum space (see Fig.\,\ref{im:SIdisp}). The splitting between $XY$ modes in both regimes is inverted.

 The fixed  energy of the excitation laser selects an approximate circle in momentum space. Traversing the $k$-space circle of excited modes results in spin rotation which is in opposite direction between the two regimes.

The real-space polarisation textures of the eigenmodes of equation~\eqref{eq:2} can be investigated at room temperature using polarisation-resolved imaging of light transmitted through the LC microcavity.
The exact polarisation state of light can be determined by a measurements of Stokes parameters $S_1$, $S_2$, $S_3$ defined as the degree of linear [$S_1$ for $X$ (horizontal) and $Y$ (vertical) linear polarisations, $S_2$ for diagonal and antidiagonal linear polarisations] and circular polarisation ($S_3$). The Stokes parameters corresponding to a second order meron, given by its analytical form [see equation~\eqref{eq:3} in Methods], are presented in Fig.\,\ref{im:Fig_Exp}a--c. The overlaid black arrows in the $S_3$ maps correspond to $\mathbf{S}_\parallel = (S_1,S_2)$. The same symmetry can be observed experimentally in the spatially resolved polarisation pattern of circularly polarised light transmitted through the LC microcavity in the $(N,N)$ regime as shown in Fig.\,\ref{im:Fig_Exp}d--f. 

Similarly, analytical pseudospin texture of a second order antimeron is depicted in Fig.\,\ref{im:Fig_Exp}g--i. As expected from numerical modelling, such polarisation texture can be observed in $(N+2,N)$ regime. Experimental results, presented in Fig.\,\ref{im:Fig_Exp}j--l, reveals second order antimeron texture. 
In the case of circularly polarised $\sigma^+$ incident light the LC microcavity, in $(N+2,N)$ regime, acts as a full-waveplate and the $\sigma^+$ light is directly transmitted, which gives a maximum for $S_3$ in the centre of the incidence spot at $x=y=0$.  Off-centre polarisation becomes linear far from the centre of the topological texture. The rotation of arrows around the centre in Fig.\,\ref{im:Fig_Exp}d,j indicates the rotation of the axis of linear polarisation. The difference between the second order meron (Fig.\,\ref{im:Fig_Exp}d) and antimeron (Fig.\,\ref{im:Fig_Exp}j) is associated with the direction of rotation of linear polarisation axis. Along a clockwise directed path around the centre of the light spot the polarisation axis rotates clockwise for a meron and anticlockwise for an antimeron. The difference is clearly visible in the real-space patterns of $S_2$ depicted in Fig.\,\ref{im:Fig_Exp} and reveals exactly the same rotation of polarisation in the reciprocal space in Fig.\,\ref{im:SIdisp}. It is straightforward to derive from equation~\eqref{eq:1} that the two opposite vorticities correspond to opposite topological integer charge $Q$. The precise size and orientation of the merons depends on the birefringence of the LC filling the cavity and the energy of the optical mode relative to the centre of the stopband (see Suppl. Inform. Fig.\,S7 and Fig.\,S8).

%\section{Discussion}
In our study we have provided the first experimental observation of a second order meron and antimeron in an electromagnetic field. The meron and antimeron polarisation textures result from the anisotropic refractive index of our optical liquid-crystal filled cavity. The artificial photonic gauge field which couples the cavity photon motion with its polarisation enables the emergence of vortical polarisation patterns. The flexibility in designing topological spin textures of light can be further combined in optical lattices mimicking magnetic order \cite{Shibata_NatNano2013} or integrated with photonics devices.
Furthermore, our findings are of fundamental interest to other systems described by models hosting analogous textures such as the Yang-Mills gauge theory or non-linear sigma models. These cavity merons can be described as a novel high order optical vector vortex state, providing a new element of structured light for study in the field of optical physics with potential application in communication, and high resolution imaging~\cite{Qiu_Science2017}. Our work opens new perspectives on using merons as topologically robust optical quaternary memory elements determined by combination of two orthogonal flows of spin (polarisation) vorticity and two opposite orientations of spin polarity.

%\bibliographystyle{naturemag}
%\bibliography{bib}

\section*{Methods}
Skyrmionic textures can be written in polar coordinates as \cite{Nagaosa_NatMater2013}:
\begin{equation} \label{eq:3}
\mathbf{S} = \left[ \cos{v \varphi} \sin{\Theta\!\left(r\right)} , \sin{v \varphi} \sin{\Theta\!\left(r\right), \cos{\Theta\!\left(r\right)}}   \right].
\end{equation}
Meron textures in Fig.\,\ref{im:Fig1}a and Fig.\,\ref{im:Fig_Exp}a--f are plotted for $\cos{\Theta\!\left(r\right)} = 0.5\left(\cos{ \pi r} + 1 \right)$, where $r\leq1$.

The polarisation of light coming from the cavity is described through the standard definition of the Stokes parameters,
\begin{align} \notag
    S_1 & = \frac{I_X - I_Y}{I_X + I_Y}, \\
    S_2 & = \frac{I_d - I_a}{I_d + I_a}, \\ \notag
    S_3 & = \frac{I_{\sigma^+} - I_{\sigma^-}}{I_{\sigma^+} + I_{\sigma^-}}.
\end{align}
Here, $I_{X,Y}, I_{d,a}, I_{\sigma^+, \sigma^-}$ correspond to the intensities of horizontal, vertical, diagonal, antidiagonal, right-hand circular and left-hand circular polarised light.

\textbf{Simulations} Berreman method\,\cite{Berreman_1972,Schubert_PRB1996} was used to calculate electric field transmitted at different incidence angles corresponding to varying in-plane wave vectors. Electric field in real space was obtained as a Fourier transform of the results in reciprocal space multiplied with a Gaussian envelope with dispersion $\sigma_x = 0.9$\,$\upmu$m in real space.

Simulations in Fig.\,\ref{im:Fig2big} are made for cavity centred at 750\,nm consisted of 8 pairs with refractive indices $n_\textrm{low} = 1.45$ and $n_\textrm{high} = 2.2$. Cavity is filled with birefringent material with $n_\textrm{o} = 1.539$ and $n_\textrm{e} = 1.939$. $(N,N)$ regime (Fig.\,\ref{im:Fig2big}a--d) is realised at long optical axis along $z$ direction and $(N+2,N)$ regime  (Fig.\,\ref{im:Fig2big}e--h) for 24.77\,deg angle between director and $z$ axis. Transmission wavelength is equal to 748.9\,nm.

\textbf{Experiment} Experimental results were obtained in a polarisation-resolved tomography measurement. Light from a broadband halogen lamp was circularly polarised and focused on a given sample with a 100$\times$ microscope objective. Transmitted light was collected by a 50$\times$ microscope objective, polarisation resolved  and focused with a 400\,mm lens on a slit of a monochromator equipped with a CCD camera. Full image was obtained by movement of the lens parallel to the slit. Experimental spatial polarisation textures presents constant energy cross sections around 10\,meV above the resonances of the cavities at normal incidence, as shown in Fig.\,S3 and Fig.\,S4. 

\textbf{$(N,N)$ sample} Experimental results presented in Fig.\,\ref{im:Fig_Exp}d--f were obtained on a cavity made of DBRs with 6 pairs of SiO\textsubscript{2}/TiO\textsubscript{2} layers designed for maximum reflectance at $\approx\!700$\,nm. $\approx2$\,$\upmu$m thick cavity is filled with birefringent liquid crystal with $n_\textrm{o} = 1.504$ and $n_\textrm{e} = 1.801$ with director oriented along $z$ direction (HT alignment). Cavity mode resonance occurs at 768.5\,nm. Transmission wavelength was equal to 763.3\,nm.

\textbf{$(N+2,N)$ sample} Experimental results presented in Fig.\,\ref{im:Fig_Exp}j--l were obtained on a cavity made of DBRs with 5 pairs of SiO\textsubscript{2}/TiO\textsubscript{2} layers designed for maximum reflectance at $\approx580$\,nm. $\approx2$\,$\upmu$m  thick cavity is filled with birefringent liquid crystal with $n_\textrm{o} = 1.539$ and $n_\textrm{e} = 1.949$ with director oriented along $x$ axis (HG alignment). Experiments were performed with square waveform with frequency 1\,kHz and peak-to-peak amplitude of 1.425\,V applied to ITO electrodes which rotates LC molecules towards $z$ axis resulting in close to degenerate cavity modes in horizontal and vertical polarisations at 583.9\,nm and 584.3\,nm. Transmission wavelength was equal to 581.5\,nm.

\textbf{Role of symmetry} The eigenvalue problem for the modes in the birefringent cavity can be
analysed from the point of view of the symmetry.  Since we are dealing
with the coupling of two modes we wish to express the relevant
Hamiltonians as second order polynomials in $k_x$ and $k_y$ with
coefficients given by combinations of Pauli matrices. In our
considerations we have to take into account the fact that the
transformation law for the Pauli matrices in each case reflects the
symmetry of the basis functions under discussion.

1) In the case of the $(N,N)$ resonance ($\epsilon_{xz}=0$) the symmetry of the system is
given by the group $D_{\infty h}$ with rotation symmetry about the $z$
axis and reflection plane perpendicular to the $z$ axis.

It easy to verify, that under the reflection in the mirror  $xy$ plane all
the Pauli matrices remain invariant while under the rotation by the angle $\phi$ about
the $z-$ axis only the $\hat\sigma_y $ matrix remains  invariant while 
$(\hat\sigma_z\pm i\hat\sigma_x) \rightarrow e^{\mp
  2i\phi}(\hat\sigma_z\pm i\hat\sigma_x)$. 
Taking into account that  under this rotation $k_x\pm ik_y
\rightarrow  e^{\mp
  i\phi}(k_x\pm ik_y)$ and that the only invariant of second order is
equal to 
$k_x^2+k_y^2$
we can postulate the following form of the Hamiltonian:
\begin{equation}\label{eqsi23}
  \begin{aligned}
\hat H &\sim \alpha_0\hat\sigma_y + \alpha_1\hat\sigma_0+
\alpha_2\hat\sigma_y(k_x^2+k_y^2)
 + \alpha_3\hat\sigma_0(k_x^2+k_y^2) +\\&
 +(\alpha_4+i\alpha_5) (\hat\sigma_z+ i\hat\sigma_x) (k_x- ik_y)^2+\\&
 +(\alpha_4-i\alpha_5) (\hat\sigma_z- i\hat\sigma_x) (k_x +ik_y)^2\\
 &\sim \alpha_0\hat\sigma_y + \alpha_1\hat\sigma_0+
\alpha_2\hat\sigma_y(k_x^2+k_y^2)
 + \alpha_3\hat\sigma_0(k_x^2+k_y^2) +\\&
 +2\alpha_4 (\hat\sigma_z(k_x^2-k_y^2)+ 2\hat\sigma_x k_kk_y)+\\&
 -2\alpha_5 (\hat\sigma_x(k_x^2-k_y^2)- 2\hat\sigma_z k_kk_y),\\
 \end{aligned}
\end{equation}
with all coefficients $\alpha_i$ - real, due to the hermiticity
requirement. Under the rotation by $\pi$ around the $x$ axis we have
$E_x\rightarrow E_x$, $E_y\rightarrow -E_y$, so $\hat\sigma_z$ remains
invariant and $\hat\sigma_x$ changes sign.  Under the same
transformation also the term $k_xk_y$ changes sign so the term
proportional to $\alpha_5$ is not invariant and we have to set
$\alpha_5=0$.  Finally, the time reversal symmetry which in this
representation is equivalent to the complex conjugation requires that
$\alpha_0=\alpha_2= 0$. If we also set
$\alpha_1=0$ we obtain the most general form of the Hamiltonian
admitted by the symmetry:
\begin{equation}\label{eqsi24}
\hat H \sim
 \alpha_3\hat\sigma_0(k_x^2+k_y^2) 
 +2\alpha_4 (\hat\sigma_z(k_x^2-k_y^2)+ \hat\sigma_xk_xk_y).
\end{equation}
with two parameters related to $\epsilon_{xx}$ and $\epsilon_{zz}$.

2) In the case of the $(N+2,N)$ resonance $\epsilon_{xz}\neq 0$ and
the relevant symmetry group is $C_{2h}$ with the twofold rotation
symmetry about the $y$-axis.  In this case $\hat\sigma_z $
is invariant under all symmetry operations while $\hat\sigma_x$ and
$\hat\sigma_y$ change sign under rotation and reflection in the $xz$-plane. The possible
invariants are therefore $\hat\sigma_0k_x^2$, $\hat\sigma_0k_y^2$,
$\hat\sigma_zk_x^2$, $\hat\sigma_zk_y^2$, $\hat\sigma_xk_xk_y$
and $\hat\sigma_yk_xk_y$. However, the last term is excluded due to
the time reversal symmetry so the most general form of the Hamiltonian
admitted by the  $C_{2h}$ symmetry for a pair of modes of the same
parity has six parameters which can be expressed in terms of $n_o,n_e,
\theta$ and mode order $N$:
\begin{equation}\label{eqsi25}
  \begin{aligned}
\hat H &\sim(\alpha_0k_x^2 + \alpha_1k_y^2)\hat\sigma_0+(\Delta E +
\alpha_2k_x^2+\alpha_3k_y^2)\hat\sigma_z+\\&+\alpha_4k_xk_y\hat\sigma_x.
\end{aligned}
\end{equation}

\section*{Acknowledgements}

This work was supported by the Ministry of Higher Education, Poland, under project ``Diamentowy Grant'': 0005/DIA/2016/45, the National Science Centre, Poland grant 2019/35/B/ST3/04147 and the Ministry of National Defense Republic of Poland Program -- Research Grant MUT Project 13--995, UK Engineering and Physical Sciences Research Council grant EP/M025330/1 on Hybrid Polaritonics, and the RFBR projects No. 20-52-12026 (jointly with DFG) and No. 20-02-00919.

\end{document}

\title{Supplementary Information: \\ Observation of second order meron polarisation textures in optical microcavities}

\author{M.\,Kr\'ol}
\affiliation{Institute of Experimental Physics, Faculty of Physics, University of Warsaw, ul.~Pasteura 5, PL-02-093 Warsaw, Poland}
\author{H.\,Sigurdsson}
\affiliation{Department of Physics and Astronomy, University of Southampton, Southampton, SO17 1BJ, United Kingdom}
\affiliation{Skolkovo Institute of Science and Technology, Moscow, Russian Federation}
\author{K.\,Rechci\'nska}
\author{P.\,Oliwa}
\author{K.\,Tyszka}
\affiliation{Institute of Experimental Physics, Faculty of Physics, University of Warsaw, ul.~Pasteura 5, PL-02-093 Warsaw, Poland}
\author{W.\,Bardyszewski}
\affiliation{Institute of Theoretical Physics, Faculty of Physics, University of Warsaw, Poland}
\author{A.\,Opala}
\affiliation{Institute of Physics, Polish Academy of Sciences, al.\,Lotnik\'{o}w 32/46, PL-02-668 Warsaw, Poland}
\author{M.\,Matuszewski}
\affiliation{Institute of Physics, Polish Academy of Sciences, al.\,Lotnik\'{o}w 32/46, PL-02-668 Warsaw, Poland}
\author{P.\,Morawiak}
\affiliation{Institute of Applied Physics, Military University of Technology, Warsaw, Poland}
\author{R.\,Mazur}
\affiliation{Institute of Applied Physics, Military University of Technology, Warsaw, Poland}
\author{W.\,Piecek}
\affiliation{Institute of Applied Physics, Military University of Technology, Warsaw, Poland}
\author{P.\,Kula}
\affiliation{Institute of Chemistry, Military University of Technology, Warsaw, Poland}
\author{P.\,G.\,Lagoudakis}
\affiliation{Skolkovo Institute of Science and Technology, Moscow, Russian Federation}
\affiliation{Department of Physics and Astronomy, University of Southampton, Southampton, SO17 1BJ, United Kingdom}
\author{B.\,Pi\k{e}tka}%\k{e}
\author{J.\,Szczytko}
\email{Jacek.Szczytko@fuw.edu.pl}
\affiliation{Institute of Experimental Physics, Faculty of Physics, University of Warsaw, ul.~Pasteura 5, PL-02-093 Warsaw, Poland}

\begin{abstract}
\end{abstract}

\maketitle

\section{Angle-resolved spectra  corresponding to Berreman simulations}

\begin{figure*}[ht]
\centering
\includegraphics[width=\textwidth]{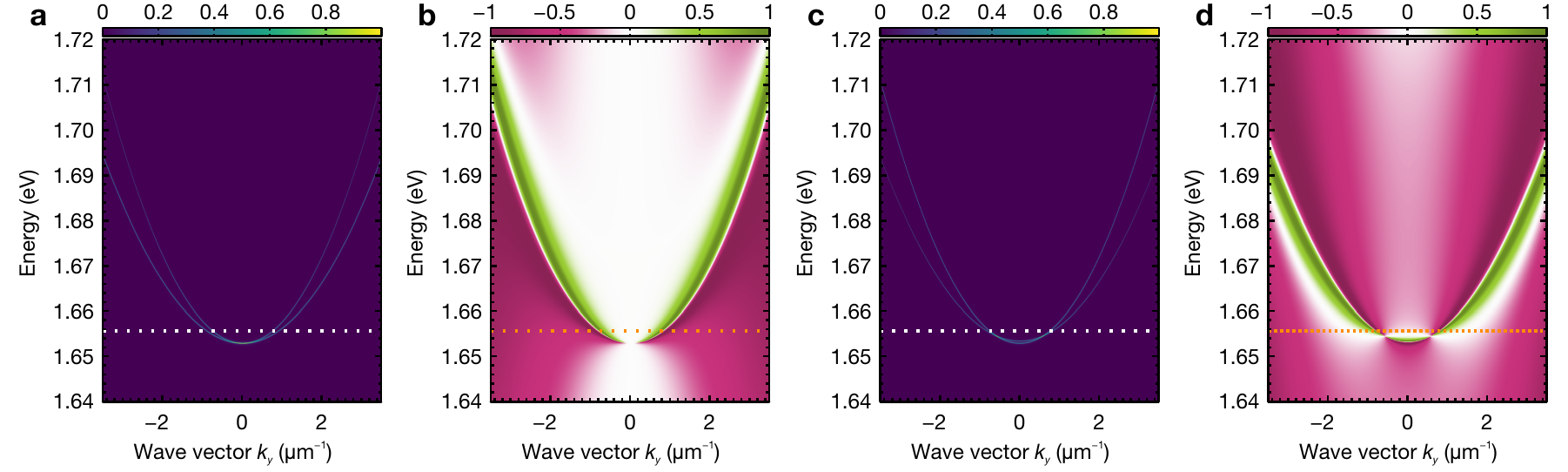}
\caption{\textbf{Simulated angle-resolved transmittance corresponding to data in Fig.\,3 in the main text.} \textbf{a} Transmittance in $(N,N)$ and \textbf{c}~$(N+2,N)$ regime. $S_1$ parameter of transmitted light in \textbf{b} $(N,N)$ and \textbf{d}~$(N+2,N)$. Dashed vertical line marks energy of transmitted light resulting in spatial polarisation textures shown in Fig.\,3 in the main tex.}
\label{im:SIFig1}
\end{figure*}

Figure\,\ref{im:SIFig1} presents simulated angle-resolved spectra corresponding to the data shown in Fig.\,3 in the main text. %Angle-resolved spectra give access to dispersion of photonic modes of a cavity since cavity mode wave vector length $k$ is connected with incidence/transmission angle $\vartheta$ by relation $k = E_\textrm{ph}/\hbar c \sin \vartheta$, where $E_\textrm{ph}$ is photon energy, $c$ is speed of light and $\hbar$ is reduced Planck constant. 
Fig.\,\ref{im:SIFig1}a shows intensity of unpolarised light transmitted through the cavity in $(N,N)$ regime ($\theta=90^\circ$, Fig.\,2a--d). We remind that $\theta$ is the angle of the liquid crystal (LC) molecular director. Fig.\,\ref{im:SIFig1}b presents corresponding $S_1$ Stokes parameter of transmitted light. Similarly Fig.\,\ref{im:SIFig1}c,d depicts transmission intensity and $S_1$ Stokes parameter for $(N+2,N)$ regime, which for this structure can be achieved by changing only molecules rotation angle to $\theta=24.77^\circ$.

% Figure\,\ref{im:NNangle}a,b presents experimental transmission intensity and $S_1$ parameter from cavity in $(N,N)$ regime where the horizontal dotted line corresponds to data shown in Fig.\,4d--f in the main text. Fig.\,\ref{im:NNangle}c,d shows simulated spectra for the same parameters as in \mbox{Fig.\,4g--i}.

% Figure\,\ref{im:NN2angle}a,b presents experimental transmission intensity and $S_1$ parameter from cavity in $(N+2,N)$ regime where the horizontal dotted line corresponds to data shown in Fig.\,4m--o in the main text. Fig.\,\ref{im:NN2angle}c,d shows simulated spectra for the same parameters as in \mbox{Fig.\,4p--r}.

% Simulations shown in Fig.\,\ref{im:SIFig1} were performed for a cavity made of mirrors with higher reflectance (simulated as higher number of DBR pairs), compared with the samples investigated experimentally (Figs.\,\ref{im:NNangle} and \ref{im:NN2angle}). Exact structures are described in Methods section of the main text. Higher reflectance of the mirrors results in narrowing of the linewidth of cavity modes. For narrow resonances modes do not overlap in momentum space at investigated photon energies, what reduces effect of rotation of the meron textures which is discussed in Section \ref{secRot}. Simulations showing orientation of second order meron texture for more realistic set of parameters when modes do overlap are presented in section\,\ref{RotSim}.

\section{Berreman matrix simulations of experimentally observed meron polarisation textures}

\begin{figure*}[ht]
\center
\includegraphics[width=\textwidth]{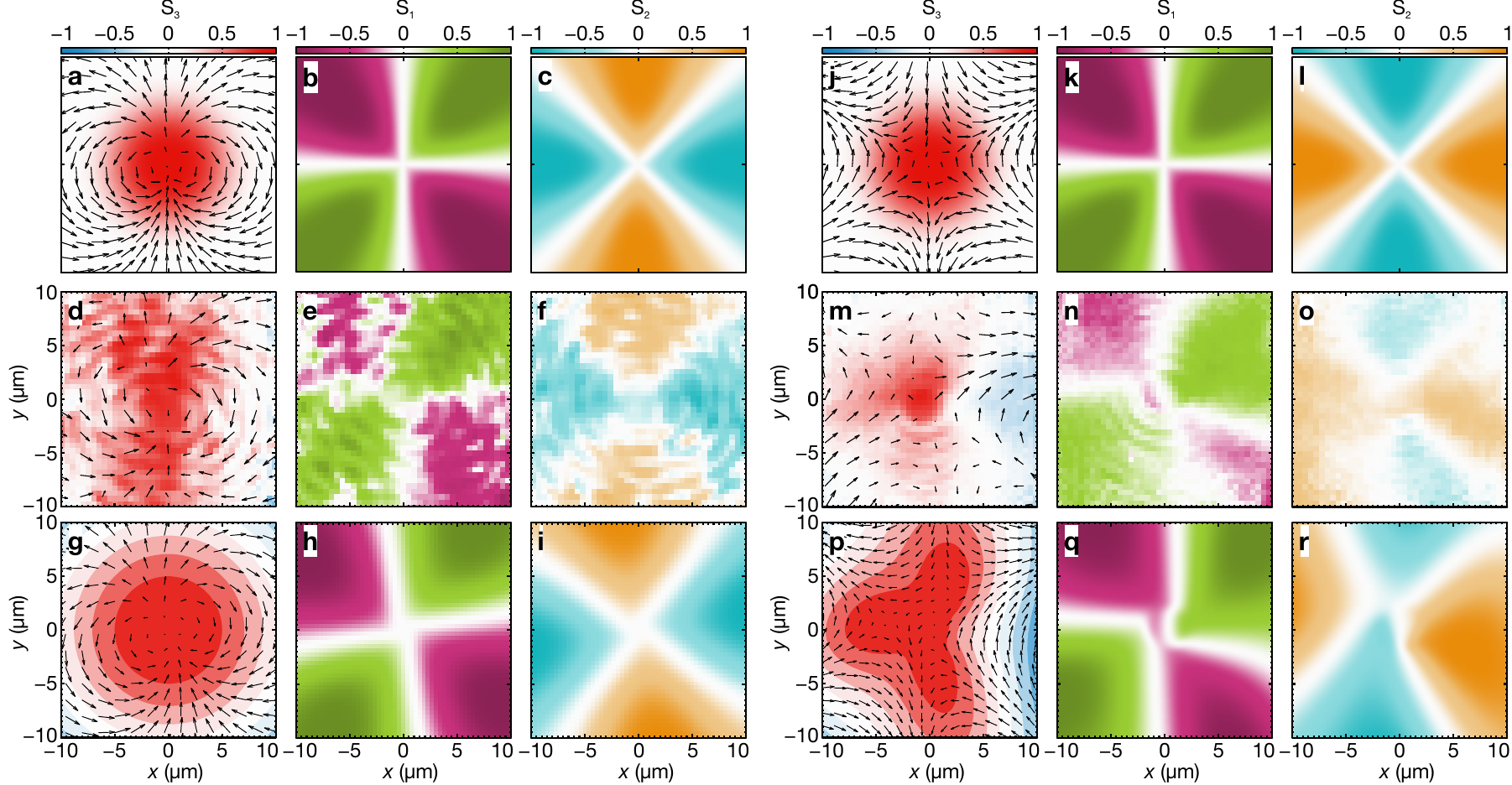}
\caption{\textbf{Second order meron and antimeron textures in LC microcavities.} \textbf{a}--\textbf{c}, $S_3, \ S_1$, and $S_2$ Stokes parameters showing the analytical spin texture of a second order meron given by equation~(3) in Methods of the main text. Black arrows correspond to $\mathbf{S}_\parallel = (S_1,S_2)$. \textbf{d}--\textbf{f}, Experimental spatial polarisation texture of $\sigma^+$ polarised light transmitted through a LC microcavity in $(N,N)$ regime. \textbf{g}--\textbf{i}, Spatial polarisation texture calculated with the Berreman method.
\textbf{j}-\textbf{l}, $S_3, \ S_1$, and $S_2$ Stokes parameters showing the analytical spin texture of a second order antimeron given by equation~(3) in Methods of the main text. \textbf{m}--\textbf{o}, Experimental spatial polarisation texture of $\sigma^+$ polarised light transmitted through a LC microcavity in $(N+2,N)$ regime. \textbf{p}--\textbf{r}, Spatial polarisation texture calculated with the Berreman method. Panels \textbf{a}--\textbf{f},\textbf{j}--\textbf{o} are a part of Fig.\,4 from the main text.}
\label{im:Fig_ExpSi}
\end{figure*}

\begin{figure*}[ht]
  \centering
      \includegraphics[width=\textwidth]{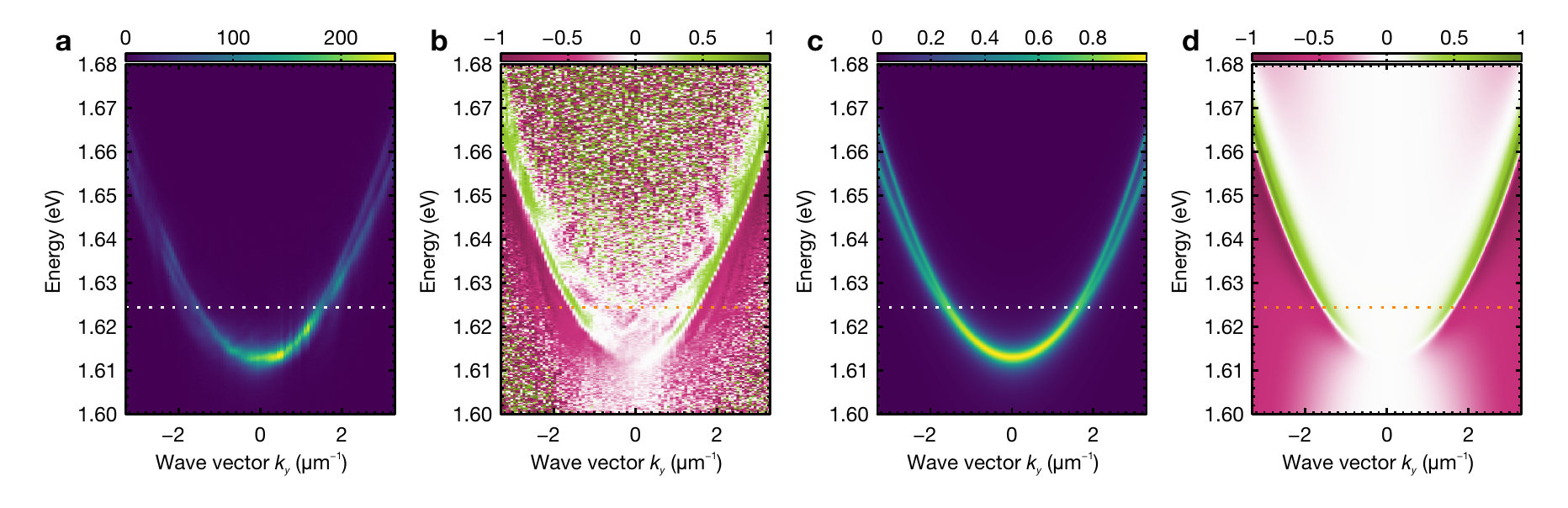}

    \caption{\textbf{Transmission from the $(N,N)$ LC microcavity.}  \textbf{a} Experimental angle-resolved transmittance of white light through $(N,N)$ LC microcavity. \textbf{b} $S_1$ stokes parameter of transmitted light. \textbf{c} Simulated angle-resolved transmittance of the cavity and \textbf{d} simulated $S_1$ Stokes parameter. Dotted vertical lines mark energy of transmitted light resulting in spatial polarisation textures shown in Fig.\,\ref{im:Fig_ExpSi}d--f corresponding to experiment and Fig.\,\ref{im:Fig_ExpSi}g--i to simulation.}%
  \label{im:NNangle} 
\end{figure*}

\begin{figure*}[ht]
  \centering
      \includegraphics[width=\textwidth]{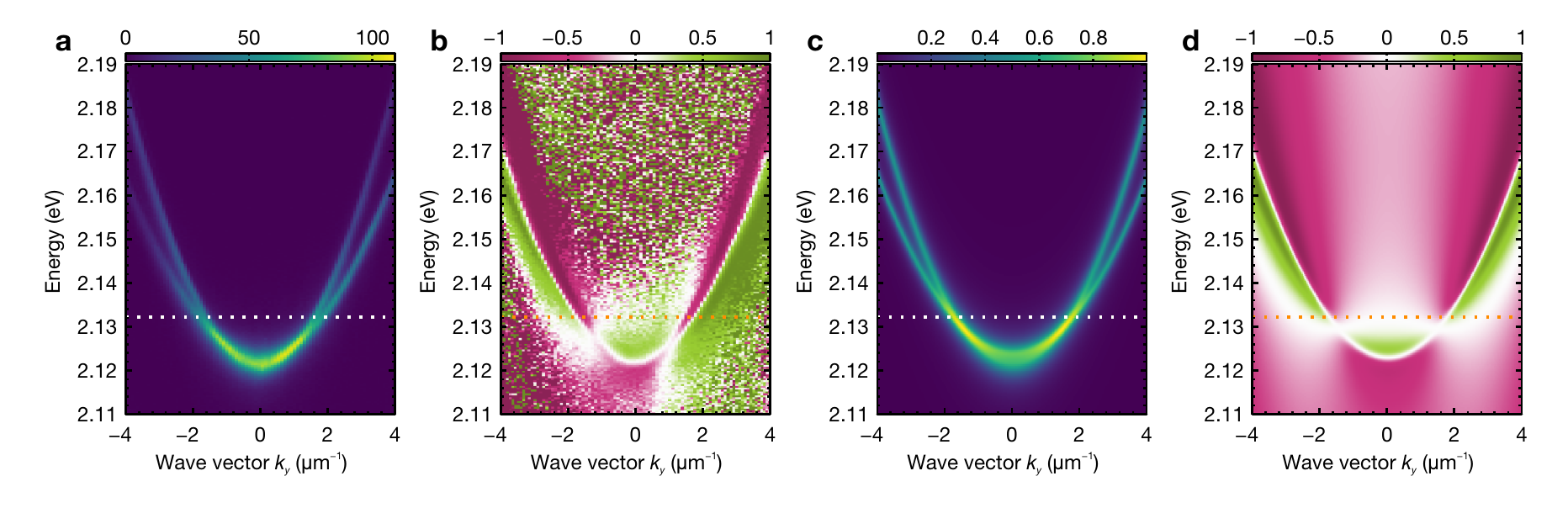}

    \caption{\textbf{Transmission from the $(N+2,N)$ LC  microcavity.}  \textbf{a} Experimental angle-resolved transmittance of white light through $(N+2,N)$ LC microcavity. \textbf{b} $S_1$ stokes parameter of transmitted light. \textbf{c} Simulated  angle-resolved transmission coefficient of the cavity and \textbf{d} simulated $S_1$ Stokes parameter. Dotted vertical lines mark energy of transmitted light resulting in spatial polarisation textures shown in Fig.\,\ref{im:Fig_ExpSi}m--o corresponding to experiment and Fig.\,\ref{im:Fig_ExpSi}p--r to simulation.}%
  \label{im:NN2angle} 
 \end{figure*}

Figure\,\ref{im:Fig_ExpSi} presents Fig.\,4 from the main text extended by Berreman matrix simulations of experimentally  observed spatial polarisation textures in $(N,N)$ regime (Fig.\,\ref{im:Fig_ExpSi}g--i) and in $(N+2,N)$ regime (Fig.\,\ref{im:Fig_ExpSi}p--r). Exact parameters of the simulated structures were optimised to match with experimental angle-resolved spectra for a given sample,  shown in Fig.\,\ref{im:NNangle} and Fig.\,\ref{im:NN2angle}.

Figure\,\ref{im:NNangle}a,b presents experimental transmission intensity and $S_1$ parameter from cavity in $(N,N)$ regime corresponding to data shown in Fig.\,\ref{im:Fig_ExpSi}d--f. Fig.\,\ref{im:NNangle}c,d shows simulated spectra for a cavity that consists of two DBRs made of 5 pairs of layers with refractive indices $n_\textrm{low} = 1.45$ and $n_\textrm{hi} = 2.2$ centred at $\lambda_0 = 700$\,nm. Simulated cavity is 1855\,nm thick  and filled with birefringent liquid crystal with $n_\textrm{o} = 1.504$ and $n_\textrm{e} = 1.801$ with director oriented along $z$ direction.

Figure\,\ref{im:NN2angle}a,b presents experimental transmission intensity and $S_1$ parameter from cavity in $(N+2,N)$ regime corresponding to data shown in Fig.\,\ref{im:Fig_ExpSi}m--o. Fig.\,\ref{im:NN2angle}c,d shows simulated spectra for a cavity that consists of two DBRs made of 4 pairs of layers with refractive indices $n_\textrm{low}$ and $n_\textrm{hi}$ centred at $\lambda_0 = 580$\,nm. 1902\,nm thick cavity is filled with birefringent liquid crystal with $n_\textrm{o} = 1.539$ and $n_\textrm{e} = 1.949$ with molecules rotation angle  $\theta = 26.27$\,deg.

\section{Coupling of cavity modes in $(N+2,N)$ regime}

%Figure\,\ref{im:SItuningDisp} presents experimental angle-resolved transmittance spectra for a cavity at varying external voltage crossing $(N+2,N)$ modes resonance probing wave vector at different directions. For wave vectors along the $x$ (Fig.\,\ref{im:SItuningDisp}a--e) and along $y$ (Fig.\,\ref{im:SItuningDisp}f--j) axes the $X$-polarised mode gradually crosses the $Y$-polarised mode. However for the antidiagonal wave vector direction ($k_x = -k_y$) an anticrossing behaviour between the modes can be observed, which is an evidence on coupling between them.

\begin{figure*}[ht]
\centering
\includegraphics[width=\textwidth]{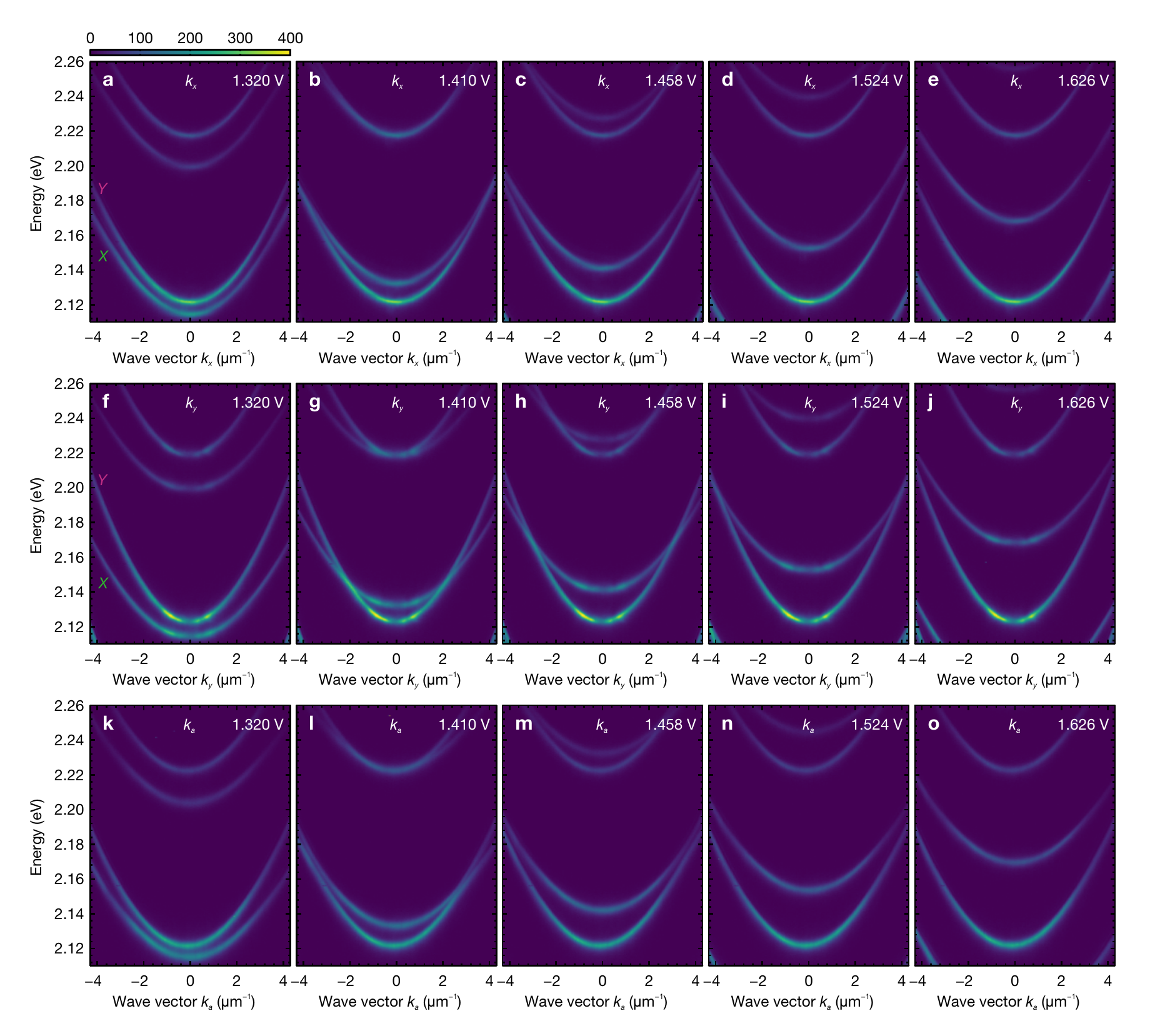}
\caption{\textbf{Angle-resolved transmission intensity at different voltages applied to the LC microcavity around the $(N+2,N)$ regime showing the gradual change in the system's dispersion properties.} \textbf{a}--\textbf{e} Transmission angle along $x$ axis:  \textbf{a}\,1.320\,V, \textbf{b}\,1.410\,V, \textbf{c}\,1.458\,V  \textbf{d}\,1.524\,V and \textbf{e}\,1.626\,V. The changing $X$ polarised mode crosses over the unaffected $Y$ polarised mode. \textbf{f}--\textbf{j}\,Transmission angle along $y$ axis:  \textbf{f}\,1.320\,V, \textbf{g}\,1.410\,V, \textbf{h}\,1.458\,V  \textbf{i}\,1.524\,V and \textbf{j}\,1.626\,V. As previously, the changing $X$ polarised mode crosses the unaffected $Y$ polarised mode. \textbf{k}--\textbf{o} Transmission angle along antidiagonal ($a$) direction ($k_x = -k_y$):  \textbf{k}\,1.320\,V, \textbf{l}\,1.410\,V, \textbf{m}\,1.458\,V  \textbf{n}\,1.524\,V and \textbf{o}\,1.626\,V. Increasing voltage now reveals the coupling between the modes observed as anticrossing behaviour.}
\label{im:SItuningDisp}
\end{figure*}

Figure\,\ref{im:SItuningDisp} presents experimental angle-resolved transmittance spectra for a cavity tuned around $(N+2,N)$ crossing (varying external voltage). Fig.\,\ref{im:SItuningDisp}a--e presents dispersion relation for wave vectors along $x$ direction, Fig.\,\ref{im:SItuningDisp}f--j along $y$ direction and Fig.\,\ref{im:SItuningDisp}k--o along diagonal direction.  For wave vectors along the $x$ and along $y$ axes the $X$-polarised mode gradually crosses the $Y$-polarised mode. However for the antidiagonal wave vector direction ($k_x = -k_y$) an anticrossing behaviour between the modes can be observed, which is an evidence on coupling between them. 

\begin{figure*}[ht]
\centering
\includegraphics[width=\textwidth]{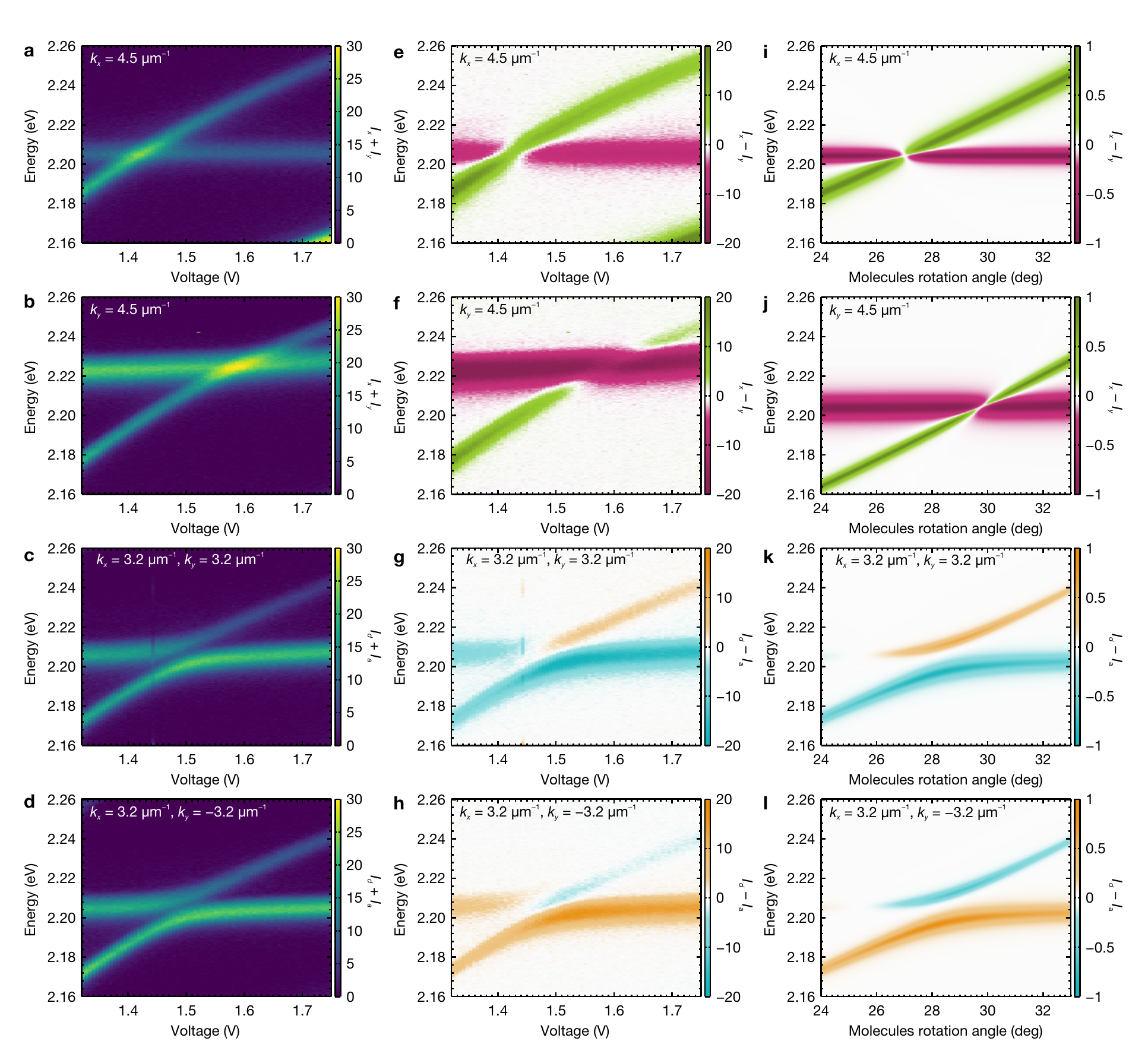}
\caption{\textbf{Voltage tuning of LC microcavity in $(N+2,N)$ regime at 4.5\,$\upmu$m$^{-1}$ wave vector at different directions.} Total transmission intensity at \textbf{a}\,$k_x = 4.5$\,$\upmu$m$^{-1}$, \textbf{b}\,$k_y = 4.5$\,$\upmu$m$^{-1}$, \textbf{c}\,$k_d = 4.5$\,$\upmu$m$^{-1}$ and \textbf{d}\,$k_a = 4.5$\,$\upmu$m$^{-1}$. \textbf{e}--\textbf{f} Difference between transmission intensities of $X$- and $Y$-polarised light corresponding to (\textbf{a},\textbf{b}). \textbf{g}--\textbf{h} Difference between transmission intensities of diagonally and antidiagonally polarised light corresponding to (\textbf{c},\textbf{d}). \textbf{i}--\textbf{l} Corresponding simulated difference between transmittance in relevant polarisations with rotation of LC molecules director.}
\label{im:SItuning}
\end{figure*}

This anticrossing can be better illustrated in Fig.\,\ref{im:SItuning}, showing transmission intensity at different voltages at a fixed  4.5\,$\upmu$m$^{-1}$ wave vector value oriented in different directions: Fig.\,\ref{im:SItuning}a for $k_x$, Fig.\,\ref{im:SItuning}b for $k_y$, Fig.\,\ref{im:SItuning}c for $k_d$ and Fig.\,\ref{im:SItuning}d for $k_a$. With wave vector along $x$ and $y$ directions are polarised accordingly to the main axes of LC molecules as shown in Fig.\,\ref{im:SItuning}e,f presenting intensity difference between $X$-polarised transmission intensity ($I_X$) and  $Y$-polarised intensity ($I_Y$). At those directions modes crosses each other. Detection along the diagonal and antidiagonal directions (Fig.\,\ref{im:SItuning}c,d) reveals coupling between the modes observable as  anticrossing behaviour. For these wave vector orientations there is significant difference between intensity detected in diagonal ($I_\textrm{d}$) and antidiagonal ($I_\textrm{a}$) linear polarisations as presented in (Fig.\,\ref{im:SItuning}g,h). Experimentally observed results are in a good agreement with Berreman matrix simulations shown in Fig.\,\ref{im:SItuning}i--l.  

\section{Meron orientation and size \label{RotSim}}

\begin{figure*}[ht]
\centering

\includegraphics{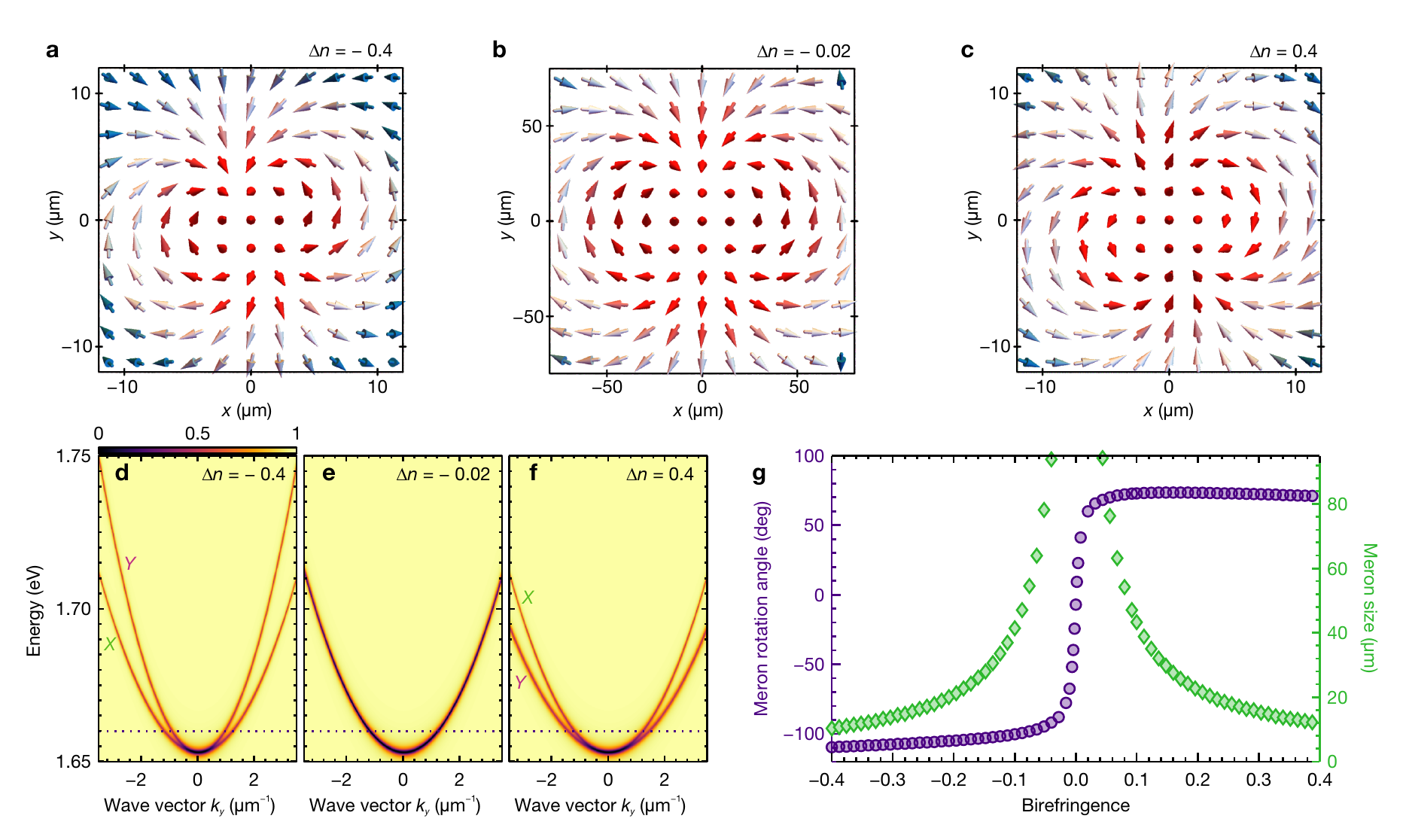}
\caption{\textbf{Simulated dependence of the second order meron orientation and size on LC birefringence.} Polarisation texture of transmitted light for \textbf{a}\,$\Delta n = -0.4$, \textbf{b}\,$\Delta n = -0.02$ and \textbf{c}\,$\Delta n = 0.4$. Note the flipped in-plane orientation of the arrows. Angle-resolved reflectance spectra for \textbf{d}\,$\Delta n = -0.4$, \textbf{e}\,$\Delta n = -0.02$ and \textbf{f}\,$\Delta n = 0.4$ where dashed line marks photon energy investigated in transmission. \textbf{g}\,Dependence of the size (radius) and orientation angle of a second order meron (diamonds and circles respectively) on LC birefringence.}
\label{im:SIrot}
\end{figure*}

Size and orientation of the meron polarisation texture depends on the exact properties of a given LC microcavity. Fig.\,\ref{im:SIrot} presents impact of the birefringence of LC layer. Berreman matrix simulations were performed for a cavity made of 5 distributed Bragg reflector (DBR) pairs of layers with refractive indices $n_\textrm{low} = 1.45$ and $n_\textrm{high} = 2.2$ and thickness $\lambda_0/4n_i$, where $\lambda_0 = 750$\,nm (1.6531\,eV). Central LC layer was simulated with $n_\textrm{o} = 1.504$ and thickness $5 \lambda_0/n_\textrm{o}$, where $n_\textrm{e}$ was changed to obtain different birefringence $\Delta n = n_\textrm{e}-n_\textrm{o}$. Fig.\,\ref{im:SIrot}a--c presents simulated spatial polarisation textures of transmitted light obtained for $\sigma^+$ polarised incident beam with wavelength 748.9\,nm (1.6556\,eV) at different birefringences:  Fig.\,\ref{im:SIrot}a $\Delta n = -0.4$, Fig.\,\ref{im:SIrot}b $\Delta n = -0.02$ and Fig.\,\ref{im:SIrot}c $\Delta n = 0.4$. Corresponding angle-resolved reflectance spectra are presented in Fig.\,\ref{im:SIrot}d--f. With varying birefringence both spatial size and orientation of the second order meron polarisation texture changes, as summarised in Fig.~\ref{im:SIrot}g. With increasing birefringence meron texture rotates clockwise with the steepest change when $\Delta n$ is close to zero. Low optical anisotropy of the LC layer results also in increasing size of the meron texture.  Due to low light intensity far away from the excitation spot simulation range is limited to $\approx \pm 100$\,$\upmu$m.

\begin{figure*}[ht]
\centering

\includegraphics{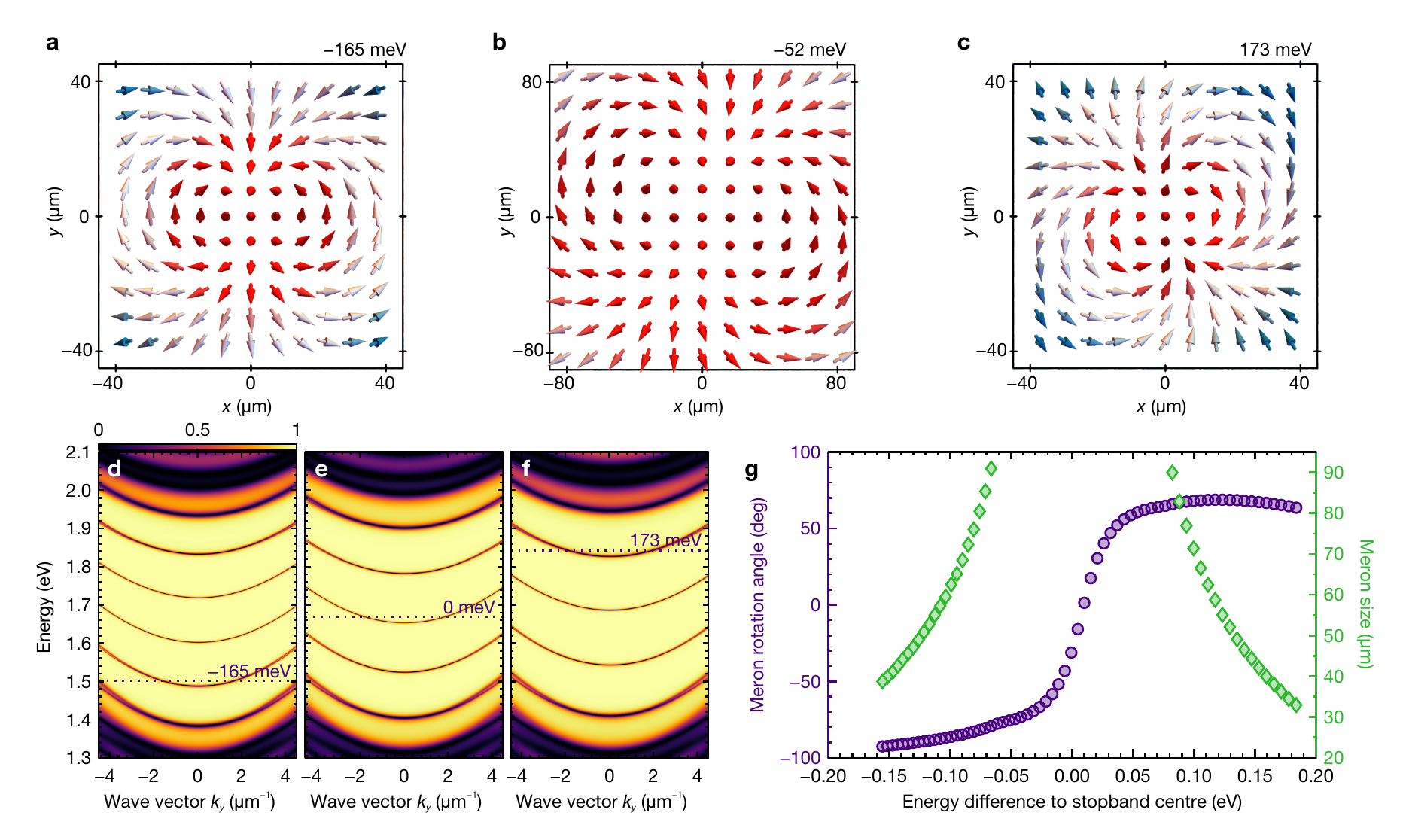}
\caption{\textbf{Simulated dependence of second order meron orientation and size on energy position of the cavity mode within photonic stopband region.} Polarisation texture of transmitted light for \textbf{a}\,$-165$\,meV, \textbf{b}\,$-52$\,meV and \textbf{c}\,$173$\,meV energy shift of the cavity mode from stopband centre. Angle-resolved reflectance spectra for \textbf{d}\,$-165$\,meV, \textbf{e}\,$0$\,meV and \textbf{f}\,$173$\,meV energy shift of the cavity mode from stopband centre, where dashed line marks photon energy investigated in transmission. \textbf{g}\,Dependence of the size (radius) and orientation angle of a second order meron (diamonds and circles respectively) on cavity mode energy shift.}
\label{im:SIrotstopband}
\end{figure*}

Size and orientation of the meron textures depends also on the energy position of the mode within the photonic stopband region of the DBRs, which is summarised in Fig.\,\ref{im:SIrotstopband}. Calculations were performed for analogous cavity as in  Fig.\,\ref{im:SIrot}, with $\Delta n =0$. Energy of the mode is changed in simulations by adjusting thickness of the LC layer filling the cavity by $-300$\,nm to $350$\,nm from initial value 2437\,nm resulting in a cavity resonance at central wavelength $\lambda_0$. Such thickness range allows to tune cavity mode energy by $\approx0.3$\,eV, as shown in the angle-resolved reflectance spectra in Fig.\,\ref{im:SIrotstopband}d for $-165$\,meV, Fig.\,\ref{im:SIrotstopband}e for $0$\,meV, and Fig.\,\ref{im:SIrotstopband}f for $173$\,meV energy shifts from $\lambda_0$. The investigated mode in this multimode cavity is marked by a dashed line showing the transmitted light energy 10\,meV above the mode resonance at normal incidence. Simulated second order antimeron textures are calculated for  Fig.\,\ref{im:SIrotstopband}a $-165$\,meV, Fig.\,\ref{im:SIrotstopband}b $-52$\,meV and Fig.\,\ref{im:SIrotstopband}c for $173$\,meV energy shift from the central wavelength. Overall dependence of meron texture orientation and size on the cavity mode energy shift (Fig.\,\ref{im:SIrotstopband}g) follows qualitatively the same dependence as when varying the birefringence shown previously in Fig.\,\ref{im:SIrot}g.

 \section{Effective Hamiltonians for coupled X and Y polarised modes}
The eigenmodes inside the cavity are represented by plane waves propagating in the plane of the cavity perpendicular to the $z$ axis:
\begin{equation}
  \label{eqsi1}
  \left(\begin{array}{l} E_x(x,y,z) \\ E_y(x,y,z)\end{array}\right) =
  \vec E_{\vec{k}} (z)e^{i( \vec{k}\cdot\vec r -\omega t)} 
\end{equation}
The vector $\vec E_{\vec{k}}$ can by found from the following effective wave equation in the birefringent medium  characterised by a dielectric tensor $\epsilon_{ij}$:
\begin{equation}\label{eqsi2}
-\partial^2_z \vec E +  \hat A  
\partial_z\vec E + \hat B_1
\vec E =  k_0^2\hat B_0
\vec E
\end{equation}
where $\vec{k} = \mathbf{k} = (k_x,k_y)$ and $k_0=\omega/c$. Assuming that $\epsilon_{xy}=\epsilon_{yx}=\epsilon_{zy}=\epsilon_{yz}=0$, we have up to the second order in $k_x$ and $k_y$:
\begin{equation}
  \label{eqsi3}
  \hat A = 
\frac{-i \epsilon_{xz}}{\epsilon_{zz}}\left[
\begin{array}{cc}
 2 k_{x} &  k_{y} \\
  k_{y} &
  0 \\
\end{array}
\right],
\end{equation}

\begin{equation}
    \label{eqsi4}
 \begin{aligned}
\hat B_1 &=
\frac{1}{\epsilon_{zz}}\left[
\begin{array}{cc}
 \epsilon_{xx} k_{x}^2+\tilde\epsilon_{zz}k_{y}^2  &
(\epsilon_{yy} -\epsilon_{zz})k_{y}k_{x} \\
(\epsilon_{xx} -\tilde\epsilon_{zz})k_{y}k_{x} &
\epsilon_{zz}k_{x}^2 +\epsilon_{yy} k_{y}^2
    \\
\end{array}\right]\\
\end{aligned}
\end{equation}
and 
\begin{equation}
  \label{eqsi5}
  \begin{aligned}
\hat B_0 &=
\left[
\begin{array}{cc}
\tilde\epsilon_{xx} &
0 \\
0 &
\epsilon_{yy} \\
\end{array}\right].\\
\end{aligned}
\end{equation}
Here, $\epsilon_{yy} = n_o^2$  and for the given angle $\theta$ between the director of the LC molecules and the $x$ axis we have

\begin{equation}
  \label{eqsi6}
  \begin{aligned}
\tilde\epsilon_{xx}
= n_{eff}^2 & = \frac{n_o^2n_e^2}{n_o^2\cos^2\theta + n_e^2\sin^2\theta} ,\\
\tilde\epsilon_{zz}  &= \frac{n_{eff}^2(n_o^4\cos^2\theta + n_e^4\sin^2\theta)}{n_o^2n_e^2},\\
\end{aligned}
\end{equation}
and $\epsilon_{xz} = \epsilon_{zx}=(n_e^2-n_o^2)\sin\theta\cos\theta$. 

We wish to find the approximate dispersion relations of modes almost
perfectly confined between the mirrors. Therefore the electric field
is expanded as follows:
\begin{equation}
  \label{eqsi7}
  \vec E_{\vec{k}}(z) = \sum_{s=X,Y}\sum_{n=1}^\infty f_{sn}|s,n\rangle,
\end{equation}
where the basis states:
\begin{equation}  \label{eqsi8}
  \begin{aligned}
    &|X,n\rangle = (-1)^n\sqrt{\frac{2}{L}}\sin\left(\frac{n\pi}{L}z\right)
\left[\begin{array}{c}1\\0\\\end{array}\right]\\
\text{and}&\\
&|Y,n\rangle =  (-1)^n\sqrt{\frac{2}{L}}\sin\left(\frac{n\pi}{L}z\right) 
\left[\begin{array}{c}0\\1\\\end{array}\right]
\end{aligned}
\end{equation}
with $n = 1,2,3\ldots$, correspond to the electric field
polarised parallel to the $x$ and $y$ axis,
respectively. In this representation the matrix elements:
\begin{equation}  \label{eqsi9}
\langle sn|\partial^2_z|s'n'\rangle = -\frac{\pi^2}{L^2}n^2\delta_{nn'}\delta_{ss'},
\end{equation}

\begin{equation}  \label{eqsi10}
\langle sn|\hat B_{1,0}|s'n'\rangle = (\hat B_{1,0})_{ss'}\delta_{nn'}.
\end{equation}
couple modes of the same order 
while the matrix elements 
\begin{equation}  \label{eqsi11}
\langle sn|\hat A \partial_z|s'n'\rangle =  (\hat A)_{ss'}
\left\{\begin{array}{cl}\dfrac{4nn'}{L(n'^2-n^2)}&\,\text{for $n'+n$ odd,}\\
0&\,\text{for $n'+n$ even}\\
\end{array}\right.
\end{equation}
couple only modes with different parity. 

At  $k_x = k_y = 0$ we have simple modal solutions  with the electric field
$\vec E_{x,n} = |X,n\rangle$ polarised along the $x$ axis with  
 frequency  $\omega_{Xn} = c k_{Xn} = c\pi n/(L n_{eff})$ and
$\vec E_{y,n} = |Y,n\rangle$  modes polarised along $y$ direction  with
 $\omega_{Yn} = c k_{Yn} = c\pi n/(L n_o)$.
The degeneracy of modes occurs when
$\omega_{Xn}\approx\omega_{Yn}\approx
\omega_0=\sqrt{(\omega_{Xn}^2+\omega_{Yn}^2)/2}$. In order to find the
approximate dispersion relation for frequencies in the vicinity of
$\omega_0$ we solve the system of linear equations for
expansion coefficients $f_{sn}$:
\begin{equation}  \label{eqsi12}
  \sum_{s'=X,Y}\sum_{n'=1}^{\infty} (\hat W)_{sn,s'n'}f_{s'n'} = 0
\end{equation}
where
\begin{equation}  \label{eqsi13}
  \begin{aligned}
    (\hat W)_{sn,s'n'} & = \left(\frac{\pi^2}{L^2}n^2\delta_{ss'} +(\hat B_1)_{ss'} -  k_0^2(\hat B_0)_{ss'}\right)\delta_{nn'} \\ &+
  \langle sn|\hat A \partial_z|s'n'\rangle.\\
  \end{aligned}
\end{equation}
In the matrix form we have:
\begin{equation}  \label{eqsi14}
  \hat W\cdot \vec f = 0.
  \end{equation}
  Note that the last term in Eq.~\eqref{eqsi13} is linear in $\vec k$
  so the coupling of modes of different parity can be treated
  perturbatively. In particular, when the degenerate modes are of the
  same parity, for example $n=n'= N$ or $n = N$ and $n' = N+2$, this
  last term will lead to the correction of the second order and higher
  in $\vec k$. In order to see this we can introduce the projection
  operator $\hat P$ on the modes of the same parity as $N$ ($\hat P$-parity), and
  $\hat Q$ - the projection operator on the modes of opposite
  parity ($\hat{Q}$-parity). Then of course
  $\vec f = \hat P\cdot \vec f + \hat Q\cdot\vec f$ where the first
  term constitutes the dominant part of $\vec f$ and the other
  represents the admixture from the states of opposite parity. Since
  we are interested mainly in the dispersion relation, we are looking
  for the solution  for the dominant part $\hat P\cdot \vec f $:
  \begin{equation} \label{eqsi15}
    (\hat P \hat W \hat P - \hat P \hat W \hat Q (\hat Q\hat W \hat
    Q)^{-1}\hat Q\hat W \hat P) \vec f = 0.
    \end{equation}
    The matrix $\hat Q\hat W \hat Q$ is limited to the subspace of states
    with $\hat{Q}$-parity  and so is its inverse $(\hat Q\hat W \hat Q)^{-1}$. To the lowest
    (zeroth) order in $\vec k$:
    \begin{equation}\label{eqsi16}
     ( (\hat Q\hat W \hat Q)^{-1})_{sn,s'n'}=
     \delta_{ss'}\delta_{nn'}\frac{1}{\frac{\pi^2}{L^2}n^2 -
       k_0^2(\hat B_0)_{ss}}.
   \end{equation}
   The matrix $\hat Q \hat W \hat P$ which couples modes of different
   parity has the form:
    \begin{equation}\label{eqsi17}
       (\hat Q \hat W \hat  P)_{sn,s'n'} = (\hat
       A)_{ss'}\frac{4nn'}{L(n'^2-n^2)}.
     \end{equation}

The
electric field in the vicinity of the degeneracy point can be approximated by:
\begin{equation}\label{eqsi18}
\vec E_{\vec{k}}(z) = f_{Xm}|X,m\rangle + f_{Yn}|Y,n\rangle,
\end{equation}
and we can consider two situations.

1) The degeneracy of two modes of the same order $m=n=N$ occurs when $n_{eff} = n_o$, i.e., when
$\epsilon_{xz} = 0$ and $\epsilon_{yy}=\epsilon_{xx}$. In this case the mode mixing term [eq.~\eqref{eqsi17}] is equal to
zero and the effective equation for the vector $\vec f =
(f_{XN},f_{YN})^T$ is
\vspace{3mm}
\begin{widetext}
\begin{equation}\label{eqsi19}
\left[
\begin{array}{cc}
(k_0^2-k_{XN}^2)\epsilon_{xx} &
0 \\
0 &
(k_0^2-k_{YN}^2)\epsilon_{xx} \\
\end{array}\right]\vec f
= 
\frac{1}{\epsilon_{zz}}\left[
\begin{array}{cc}
 \epsilon_{xx} k_{x}^2+\epsilon_{zz}k_{y}^2  &
(\epsilon_{xx} -\epsilon_{zz})k_{y}k_{x} \\
(\epsilon_{xx} -\epsilon_{zz})k_{y}k_{x} &
\epsilon_{zz}k_{x}^2 +\epsilon_{xx} k_{y}^2
\end{array}\right]
\vec f.
\end{equation}

2) In the case of degeneracy of two modes of different order but the
same parity the mixing term is different from zero so the effective
equation for  $\vec f =
(f_{XN+2},f_{YN})^T$:
  \begin{equation}  \label{eqsi20}
    \begin{aligned}
 &\sum_{n'=N+2,N}\sum_{s'=X,Y}\left(
   \left(\frac{\pi^2}{L^2}n^2\delta_{ss'} +(\hat B_1)_{ss'} -
     k_0^2(\hat B_0)_{ss'}\right)\delta_{nn'} \right. +\\ &\left.-\sum_{m''}^\infty{}^{'}\sum_{s"=X,Y} \frac{16 n n'
      m''^2(\hat A)_{ss"}(\hat
      A)_{s"s'}}{(n^2-m''^2)(m''^2-n'^2)(\pi^2m''^2- L^2k_0^2(\hat
      B_0)_{s''s''})}\right)f_{s'n'}=0.
  \end{aligned}
\end{equation}
where the prime over summation sign means that only $m''$ with parity
different from the parity of $n$ and $n'$ which is the same as the
parity of  $N$ are included. In this way the denominator is always
different form zero. 
Approximating $ k_0^2(\hat B_0)_{ss'} \approx \pi^2(n^2+n'^2)/(2L^2)$
in the denominator of the last term and defining
\begin{equation}  \label{eqsi21}
  Z_{n,n'} = Z_{n',n} =
  \sum_{m''}^\infty{}^{'} \frac{16 n n'
    m''^2}{\pi^2(n^2-m''^2)(m''^2-n'^2)(m''^2- (n^2+n'^2)/2)}
\end{equation}
we obtain the following equation for $\vec f$ in the case of  the resonance of
modes of the order $N+2$ and $N$:
\begin{equation} \label{eqsi22}
  \begin{aligned}
&\left[
\begin{array}{cc}
(k_0^2-k_{XN+2}^2)\tilde\epsilon_{xx} &
0 \\
0 &
(k_0^2-k_{YN}^2)\epsilon_{yy} \\
\end{array}\right]\vec f = \\
&= 
\frac{1}{\epsilon_{zz}}\left[
\begin{array}{cc}
( \epsilon_{xx}+4Z_{N+2,N+2}\dfrac{\epsilon_{xz}^2}{\epsilon_{zz}})
  k_{x}^2
  +(\tilde\epsilon_{zz}+Z_{N+2,N+2}\dfrac{\epsilon_{xz}^2}{\epsilon_{zz}})k_{y}^2  &
2Z_{N+2,N}\dfrac{\epsilon_{xz}^2}{\epsilon_{zz}}k_xk_y\\
2Z_{N+2,N}\dfrac{\epsilon_{xz}^2}{\epsilon_{zz}}k_xk_y&
\epsilon_{zz}k_{x}^2 +(\epsilon_{yy}+Z_{N,N}\dfrac{\epsilon_{xz}^2}{\epsilon_{zz}})k_{y}^2
\end{array}\right]
\vec f.
\end{aligned}
\end{equation}
\end{widetext}
Note that the effective equations in the vicinity of the resonance of modes of
the same order $(N,N)$ [eq. (\ref{eqsi19})] and for the case of
different orders, $(N,N+2)$ [eq. (\ref{eqsi22})] have similar
structure. However the origin of the term proportional to $k_xk_y$,
which is responsible for coupling between the modes is different in each
situation. In the $(N,N)$ case we have a direct coupling between the TE
and TM modes whereas the coupling between modes of different order
 is of indirect character and is mediated by modes with opposite
parity. By standard manipulations, both equations can be transformed
into
an eigenvalue problem with a Hamiltonian presented in the main text.

\section{Spin structure and meron orientation from momentum-space Hamiltonian \label{secRot}}

\begin{figure*}[ht]
\centering

\includegraphics[width=10cm]{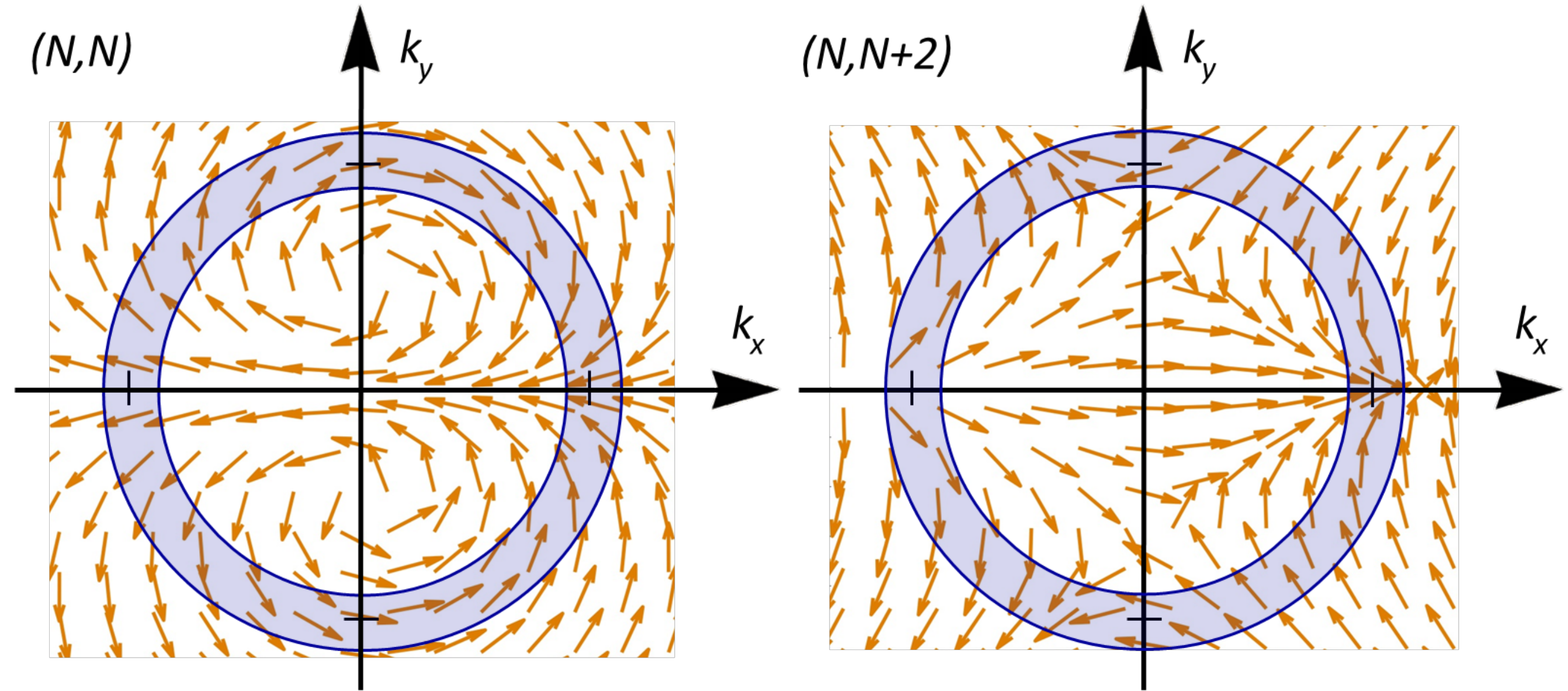}
\caption{\textbf{Spin polarization from momentum-space Hamiltonian.} The left and right panels show the spin polarization of one of the two eigenmodes of hamiltonian (2) in the main text (yellow arrows) in the meron and antimeron cases. The other mode has opposite polarization. The shaded ring depicts the approximate area in momentum space excited by a resonant laser beam. The polarization on the ring corresponds to the spin rotation in Fig.~3 in the main text.}
\label{im:SIFigteo1}
\end{figure*}

The meron and antimeron spin structure results from transmission of light through cavity modes, which can be approximately described with Hamiltonian (2) in the main text. The emergence of such structures and the topological charge $Q$ can be predicted from the eigenmodes of the Hamiltonian taking into account that the system is excited with resonant laser light with a Gaussian envelope in space. In Fig.~\ref{im:SIFigteo1} we show the spin polarisation of one of the the Hamiltonian eigenmodes in the meron $(N,N)$ and antimeron $(N+2,N)$ case. The shaded ring in momentum space corresponds to the approximate area excited with resonant light, which results from the parabolic dispersion relation of the cavity (see Fig.~2 in the main text). The second order meron spin structure of can be observed on this ring, and is retained after performing Fourier transform into real space, assuming that the excitation laser beam is Gaussian-shaped.

This simple explanation, however, is incomplete as it neglects the second, orthogonal eigenmode and does not explain the meron rotation angle discussed in the previous section. To take into account the second mode, we estimate the amplitude and polarisation of light transmitted through microcavity. The amplitude of input light can be written as

\begin{equation}
\textbf{A}_{\rm in}(\textbf{k},\omega) = A(\textbf{k}) A(\omega) \mathbf{u}_{\rm in},
\end{equation}
where $A(\textbf{k})$ is a Gaussian shaped amplitude, $A(\omega)$ is approximately $\delta$-shaped laser frequency spectrum, $\textbf{u}_{in}$ is the polarisation of input light, e.g.~in linear polarization basis $\textbf{u}_{in}=(1,0)^T$ for a horizontally polarised light. In the considered cases $(N,N)$ and $(N+2,N)$ the cavity acts as a full-wave plate, so the polarisation of cavity mode at the output is not rotated by the cavity. The output amplitude is
\begin{equation}
    \textbf{A}_{\rm out}(\textbf{k})= \sum_{i=1,2} t_i(\textbf{k}) A(\textbf{k}) P(\textbf{u}_{\rm in},\textbf{u}_i) \textbf{u}_i,
\end{equation}
where we approximate the cavity transmission coefficient $t$ as a sum of two eigenmodes $i=1,2$, each corresponding to a peak in transmission $t_i(\textbf{k})$ with a similar amplitude and a Gaussian shape. The operator $P(\textbf{u}_{\rm in},\textbf{u}_i)= \textbf{u}_{\rm in} \cdot \textbf{u}_i$ is the projection of input light polarisation on the eigenmode of the Hamiltonian (2) polarisation. The shape of $t_i(\textbf{k})$ in momentum space is ring-like for each mode, with slightly differing radii. This results from the parabolic dispersion relation of the in-plane photonic cavity modes as shown in Fig.~3 in the main text. 

\begin{figure*}[ht]
\centering

\includegraphics[width=13cm]{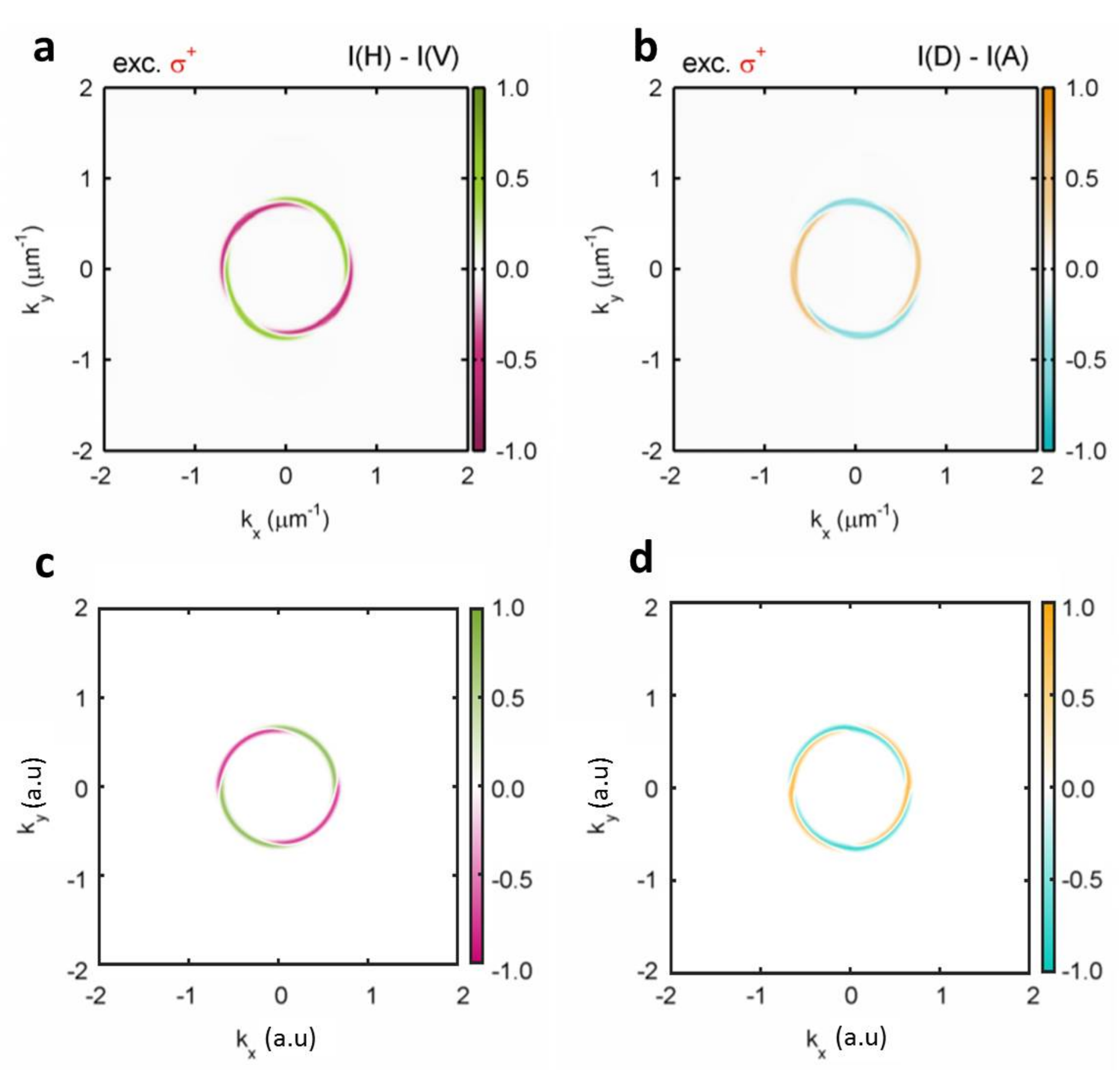}
\caption{\textbf{Polarisation of transmitted light in momentum space.} The top panels show the results of the Berreman method and the bottom panels the approximate Hamiltonian (2) in the case of circular input polarisation. The mixing of two modes with orthogonal polarisations (corresponding to rings with slightly different radii) results in rotation of the input polarisation in the diagonal directions ($k_x=\pm k_y)$, which transforms the input circular polarisation into horizontal or vertical one between the rings. This leads to a helical structure of modes visible both in X-Y (left) and A-D polarisation patterns and the rotation of orientation by approximately 45 degrees.}
\label{im:SIFigteo2}
\end{figure*}

Calculations of the above simplified Hamiltonian model are compared with Berreman method simulations in the case of $(N+2,N)$ antimeron with $\sigma^+$ excitation in Fig.~\ref{im:SIFigteo2}. The approximate 45 degrees orientation of the antimeron results from the overlap of the two rings in momentum space, with the phase of the transmission coefficients $t_i$ differing by $\pi/2$. Such phase difference is explained by the dependence of the phase of the transmission coefficient on transverse momentum. This additional phase shift leads to rotation of input circular polarisation into horizontal or vertical polarisation in the diagonal directions ($k_x=\pm k_y$), which results in the whirling polarisation structure in momentum space and the corresponding rotation of the meron orientation.

\newpage

